\documentclass[12pt]{article}

\usepackage{amssymb}
\usepackage{amsfonts}
\usepackage{amsmath}
\usepackage{bbm}
\usepackage{graphicx}
\usepackage{color}
\usepackage{rotating}
\usepackage{lscape}
\usepackage{lscape}
\usepackage{wasysym}
\usepackage{hyperref}
\usepackage{mathtools}
\hypersetup{
 colorlinks=true,
 linkcolor=blue,
 citecolor=magenta,
}

\def\Z{\mathbb Z}
\def\C{\mathbb C}
\def\R{\mathbb R}
\def\NN{\mathbb N}

\newcommand{\al}{\alpha}
\newcommand{\be}{\beta}

\newcommand{\ga}{\gamma}

\newcommand{\eps}{\epsilon}

\newcommand{\si}{\sigma}

\newcommand{\De}{{\Delta\vphantom{\big|}}}

\newcommand{\sfrac}[2]{{\textstyle\frac{#1}{#2}}}

\newcommand{\pa}{\partial}

\newcommand{\im}{{\mathrm{i}}}
\newcommand{\ep}{{\mathrm{e}}}
\newcommand{\ph}{\phantom{-}}

\newcommand{\ab}{\bar{\alpha}}
\newcommand{\bb}{\bar{\beta}}
\newcommand{\gb}{\bar{\gamma}}
\newcommand{\gl}{g_{\textrm{L}}}
\newcommand{\gs}{g_{\textrm{S}}}
\newcommand{\gls}{{g_{\textrm{L}}^2}}
\newcommand{\gss}{{g_{\textrm{S}}^2}}
\newcommand{\glc}{{g_{\textrm{L}}^3}}

\newcommand{\glq}{{g_{\textrm{L}}^4}}
\newcommand{\gsq}{{g_{\textrm{S}}^4}}
\newcommand{\eb}{{\bar{\epsilon}}}
\newcommand{\ebe}{{\epsilon\bar{\epsilon}}}

\newcommand{\beq}{\begin{equation}}
\newcommand{\eeq}{\end{equation}}
\newcommand{\eq}{\end{equation}}
\newcommand{\bea}{\begin{eqnarray}}
\newcommand{\eea}{\end{eqnarray}}
\newcommand{\with}{{\quad{\rm with}\quad}}
\newcommand{\for}{{\quad{\rm for}\quad}}
\renewcommand{\and}{{\quad{\rm and}\quad}}
\newcommand{\und}{{\qquad{\rm and}\qquad}}
\renewcommand{\=}{\ =\ }

\advance \textheight by \topskip
\makeatletter
\@addtoreset{equation}{section}

\makeatother

\begin{document}

\begin{titlepage}
\setcounter{page}{0}

\phantom{.}
\vskip 1.5cm

\begin{center}

{\LARGE\bf
${\cal PT}$ deformation of \\[12pt] Calogero--Sutherland models
}

\vspace{12mm}
{\Large Francisco Correa${}^{+}$ and \
Olaf Lechtenfeld${}^{\times}$ 
}
\\[8mm]

\noindent ${}^+${\em 
Instituto de Ciencias F\'isicas y Matem\'aticas\\
Universidad Austral de Chile, Casilla 567, Valdivia, Chile}\\
{Email: francisco.correa@uach.cl}\\[6mm]

\noindent ${}^\times${\em
Institut f\"ur Theoretische Physik and Riemann Center for Geometry and Physics\\
Leibniz Universit\"at Hannover \\
Appelstra\ss{}e 2, 30167 Hannover, Germany }\\
{Email: olaf.lechtenfeld@itp.uni-hannover.de}\\[6mm]

\vspace{12mm}

\begin{abstract}
\noindent
Calogero--Sutherland models of $N$ identical particles on a circle
are deformed away from hermiticity but retaining a $\cal PT$ symmetry. 
The interaction potential gets completely regularized, which adds to the 
energy spectrum an infinite tower of previously non-normalizable states. 
For integral values of the coupling, extra degeneracy occurs and a 
nonlinear conserved supersymmetry charge enlarges the ring of Liouville 
charges. The integrability structure is maintained. We discuss the 
$A_{N-1}$-type models in general and work out details for the cases of 
$A_2$ and~$G_2$.
\end{abstract}

\end{center}

\end{titlepage}

\section{Introduction and summary}

\noindent
Recently, integrable systems have been subjected more intensely 
to non-hermitian deformations, as has been reviewed in~\cite{Fring12-rev}. 
Specifically, ${\cal PT}$ deformations of rational Calogero models and 
their spherical reductions have been analyzed in some detail
\cite{FrZn08,fringsmith1,fringsmith2,fringsmith3,CoLe17}.
It was found that the mathematical structures and tools pertaining to
integrability are compatible with ${\cal PT}$ deformations, as long as
the latter respects the Coxeter reflection symmetries of the models.
A bonus of certain $\cal PT$ deformations is the complete regularization
of the coincident-particle singularities of Calogero models, which leads
to an enhancement of the Hilbert space of physical states by previously
non-normalizable wave functions. For integral values of the Calogero
coupling(s), most of the new states (called `odd') are energy-degenerate 
with old ones (called `even'), giving rise to a $\Z_2$-grading and
a conserved intertwiner~$Q$ on top of the Liouville integrals of motion. 
This $Q$ is often called a `nonlinear supersymmetry' charge, 
and it enhances the symmetry of the superintegrable system to a 
$\Z_2$-graded one~\cite{CoLePl13}.

Here, we carry our analysis of $\cal PT$ deformed spherically reduced 
(or angular) Calogero models~\cite{CoLe17} over to the trigonometric 
or Calogero--Sutherland case.
These models were completely solved more than 20 years ago
\cite{LaVi95,PeRaZa98,GaLoPe01}
and describe $N$ interacting
identical particles on a circle or, equivalently, one particle moving on
an $N$-torus subject to a particular external potential. The latter has
inverse-square singularities on the hyperplanes corresponding to
incident-particle locations. We shall see that, also here, there exists 
a deformation which renders the potential nonsingular and retains
the integrable structure, adding an infinite tower of new states to the
energy spectrum and allowing for a nonlinear supersymmetry-operator
mapping between `new' and `old' states. Employing $\cal PT$-deformed
Dunkl operators, we construct the deformed intertwiners (shift operators 
increasing the coupling by unity) and analyze their action on the deformed
energy eigenstates. We then find a set of deformed Liouville charges 
which intertwine homogeneously (like the Hamiltonian), so that their 
eigenvalues are preserved by the shift. Details are worked out for the 
three-body cases based on the $A_2$ and $G_2$ groups. 

Our analysis can straightforwardly be extended to any (higher-rank)
Coxeter group, but explicit expressions quickly become rather lengthy.
While Dunkl operators and Liouville charges are known for all models
(and $\cal PT$-deformed effortlessly), we are not aware of a general
classification of Weyl {\sl anti-\/}invariant\footnote{
meaning antisymmetric under any Weyl reflection}
polynomials, which will be
needed to extend the ring of Liouville charges to include all intertwiners.
Also interesting is the exploration of the hyperbolic models and, finally,
the elliptic ones.
We have employed the most simple $\cal PT$ deformation compatible
with the symmetries of the model, but there exist other options. The two
we shortly discuss in the context of the $A_2$ model do not fully 
regularize the potential, but there may be other ones more suitable.
A classification will be most welcome.

The paper is organized as follows.
After defining the $A_{N-1}$ Calogero--Sutherland model and introducing
a suitable $\cal PT$ deformation in Section~2, we describe the energy
spectrum including the eigenstates in Section~3. The following Section~4
is devoted to the construction of deformed conserved charges and 
intertwiners with the help of Dunkl operators. Sections 5 and~6 work
out the details for the $A_2$ and $G_2$ cases, respectively. 
Explicit low-lying wave functions are listed in Appendices C and~D.

\section{${\cal PT}$ deformation of Calogero--Sutherland models}

\noindent
The $N$-particle model is governed by a rank-$N$ Lie algebra~$\mathfrak{g}$. 
Translation invariance implies that $\mathfrak{g}=A_1\oplus\mathfrak{g}^\perp$,
where the $A_1$ part represents the center of mass. It may be decoupled,
but we retain it for the time being.

For $A$-type models, $\mathfrak{g}^\perp=A_{N-1}$.
They describe $N$ identical particles on a circle of circumference~$L$,
mutually interacting via a repulsive inverse-square two-body potential.
We label the $L$-periodic particle coordinates as
\beq
q_i \in \R/L\Z \with i=1,2,\ldots,N\ , \quad 
\textrm{i.e.}\quad q_i \simeq q_i+n_iL \for n_i\in\Z\ ,
\eeq
but it proves more convenient to pass to multiplicative coordinates
\beq
x_i \= \ep^{2\im\pi q_i/L}\ ,
\eeq
with the useful relations
\beq
\frac{\pa}{\pa q_i}=\frac{2\im\pi}{L}\,\frac{x_i\pa}{\pa x_i}\ ,\quad
\sin\sfrac{\pi}{L}(q_i{-}q_j) = \frac{1}{2\im}\,\frac{x_i{-}x_j}{\sqrt{x_i\,x_j}}\ ,\quad
\cot\sfrac{\pi}{L}(q_i{-}q_j) = \im\,\frac{x_i{+}x_j}{x_i{-}x_j}\ .
\eeq

The $A_{N-1}$ Calogero--Sutherland model is defined by the Hamiltonian
\beq \label{HamN}
\begin{aligned}
H(g)&\=-\frac{1}{2}\sum^N_{i=1} \frac{\pa}{\pa q_i^2}
+\frac{\pi^2}{L^2}\sum_{i<j}\frac{g(g{-}1)}{\sin^2\!\frac{\pi}{L}(q_i{-}q_j)} \\[4pt]
&\= \frac{\pi^2}{L^2}\biggl[ 2\sum^N_{i=1} \Bigl(\frac{x_i\pa}{\pa x_i}\Bigr)^2
-4\,g(g{-}1)\sum_{i<j}\frac{x_i\,x_j}{(x_i{-}x_j)^2}\biggr]\ .
\end{aligned}
\eeq
We remark on the invariance under $g\mapsto 1{-}g$.
Rather than an $N$-body problem on a circle, 
this system may also be interpreted as a single particle moving on an $N$-torus
$T^N$ and subject to a particular external potential. 
The latter's singularities on the walls of the Weyl alcove restrict the particle motion 
to a fundamental domain in the $A_1\oplus A_{N-1}$ weight space.
For later use we introduce the shorthand notation
\beq
\pa_i = \frac{\pa}{\pa x_i} \und x_{ij} = x_i{-}x_j
\eeq
as well as the totally antisymmetric degree-zero homogeneous rational function
\beq
\De \= \prod_{i<j} x_{ij} \prod_k x_k^{-\frac12(N-1)}\ . 
\eeq

To establish ${\cal PT}$ symmetry, it is necessary to identify two involutions,
a unitary $\cal P$ and an anti-unitary~$\cal T$, 
such that the deformed Hamiltonian is invariant under their product.
While for the latter we take the standard choice of complex conjugation,
the former leaves various possibilities. 
In this paper we shall choose $\cal P$ to be parity flip of all coordinates,\footnote{
For $N{=}3$, we shall ponder on some other choices later on.}
thus
\beq \label{PT1}
{\cal P} : \ q_i \mapsto -q_i \quad\Leftrightarrow\quad x_i \mapsto x_i^{-1} \und
{\cal T} : \ \im \mapsto -\im \quad\Leftrightarrow\quad x_i \mapsto x_i^{-1} \ .
\eeq

The Hamiltonian (\ref{HamN}) is parity symmetric, so a $\cal PT$-symmetric way 
of deforming can be induced by a $\cal PT$-covariant complex coordinate change.
The obvious option is
\beq \label{PTshift}
q_i\ \mapsto\ q_i^\eps = q_i+\im\eps_i \qquad\Leftrightarrow\qquad
x_i\ \mapsto\ x_i^\eps = x_i\,\ep^{-2\eps_i}
\quad\with \eps_i>0\ ,
\eeq
implying
\beq
{\cal P} : \ q_i^\eps \mapsto -q_i^{-\eps} \quad\Leftrightarrow\quad 
x_i^\eps \mapsto (x_i^{-\eps})^{-1} \und
{\cal T} : \ \im \mapsto -\im \quad\Leftrightarrow\quad 
x_i^\eps \mapsto (x_i^{-\eps})^{-1}\ . 
\eeq
Thus, the multiplicative coordinate $x_i^\eps$ is $\cal PT$ invariant for
any value of~$\{\eps_i\}$. If we do not want to deform the center-of-mass
degree of freedom we must impose the restriction~$\sum_i \eps_i=0$.

This deformation generically removes the singularities in the potential because
\beq
\frac{x_i^\eps-x_j^\eps}{\sqrt{x_i^\eps\,x_j^\eps}} \= 
\sqrt{\frac{x_i}{x_j}}\,\ep^{-\eps_{ij}} -
\sqrt{\frac{x_j}{x_i}}\,\ep^{+\eps_{ij}}
\und
\frac{x_i^\eps-x_j^\eps}{x_i^\eps+x_j^\eps} \=
\frac{x_i\,\ep^{-\eps_{ij}}-x_j\,\ep^{+\eps_{ij}}}{x_i\,\ep^{-\eps_{ij}}+x_j\,\ep^{+\eps_{ij}}}
\eeq
never vanish for $\eps_{ij}\equiv\eps_i{-}\eps_j$ different from zero.
Therefore, the deformed Hamiltonian
\beq
H^\eps\!(g) \= \frac{\pi^2}{L^2}\biggl[ 2\sum^N_{i=1} (x_i\pa_i)^2
-4\,g(g{-}1)\sum_{i<j}\frac{x_i^\eps\,x_j^\eps}{(x_{ij}^\eps)^2}\biggr]
\eeq
no longer restricts the particle motion to a single Weyl alcove
but allows it to range over the entire~$T^N$.
This space still being compact, the energy spectrum will remain discrete.
Only in the $L\to\infty$ limit we recover the rational Calogero model
with its continous spectrum. 
In the following, we drop the superscript `$\eps$'
but understand to have a generic deformation turned on with $\eps_{ij}\neq0$.

\section{The energy spectrum}

So far, the $\cal PT$ deformation (\ref{PTshift}) is fully compatible with the
integrability of the $A$-type Calogero--Sutherland model. It merely hides in the 
substitution $x_i\mapsto x_i^\eps$. This remains true for the energy spectrum:
the known energy levels are unchanged under the deformation, and the eigenstates 
are obtained from the undeformed ones simply by again deforming the coordinates. 
However, due to the disappearance of the singularities in the potential, 
previously non-normalizable eigenstates become physical, adding extra states
to the spectrum!

One popular way to completely label the energy eigenstates is by an $N$-tupel 
\beq \label{norder}
\vec n = (n_1,n_2,\ldots,n_N) \quad\with n_1\ge n_2\ge\ldots\ge n_N \ge0
\eeq 
of quasiparticle excitation numbers. 
After removing the center-of-mass energy by boosting to its rest system 
one obtains
\beq \label{E1}
E_{\vec n}(g) \= \frac{\pi^2}{L^2}\biggl[
2\sum_k n_k^2 - \sfrac2N\Bigl(\sum_k n_k\Bigr)^2 +
2g \sum_k (N{+}1{-}2k)\,n_k + \sfrac16 N(N{-}1)(N{+}1)\,g^2 \biggr]\ .
\eeq
Due to translation invariance, this expression is invariant under a common
shift $n_k\to n_k+c$. In order to remove this redundancy, we put $n_N=0$,
so that the sums over $k$ run from 1 to~$N{-}1$ only. 
The energy is bounded from below, with the ground-state value
\beq
E_0 \ \equiv\  E_{\vec 0} \= \sfrac16N(N{-}1)(N{+}1)\,g^2\sfrac{\pi^2}{L^2}
\quad\for g\ge0
\eeq
but a different lower bound (minimally zero) for $g<0$.

To study the degeneracy, we rewrite (\ref{E1}) as a sum of squares,
\beq\label{E2}
\begin{aligned}
E_{\vec n}(g) &\= \frac{\pi^2}{L^2}\sum_{k=1}^{N-1} \sfrac2{k(k+1)} \Bigl[
n_1+n_2+n_3+\ldots+n_{k-1}-k\,n_k - \bigl(N{-}\sfrac12k(k{+}1)\bigr)\,g\Bigr]^2 \\[4pt]
&\=\frac{\pi^2}{L^2}\sum_{k=1}^{N-1} \bigl[ \lambda_k - \mu_k\,g\bigr]^2
\quad\with
\end{aligned}
\eeq
\beq
\lambda_k \= \frac{n_1+n_2+n_3+\ldots+n_{k-1}-k\,n_k}{\sqrt{k(k{+}1)/2}}
\und \mu_k = \frac{N-k(k{+}1)/2}{\sqrt{k(k{+}1)/2}}\ .
\eeq
Any collection $\vec n$ of quantum numbers uniquely yields an element
$\vec\lambda=(\lambda_1,\lambda_2,\ldots,\lambda_{N-1})$ in a particular
Weyl chamber of the $A_{N-1}$ weight space~$\Lambda_{N-1}$, 
and the energy of the corresponding state is given by the radius-squared 
of a circle in~$\Lambda_{N-1}$ centered at $\vec\mu\,g$.
Since $\vec\mu$ lies outside the Weyl chamber in question, for positive~$g$ 
the minimal distance from the circle center to the physical states is given 
by $|\vec\mu|$ and represents the nonzero ground-state energy~$E_0(g)$.
So the degeneracy of a given energy level may be found by counting the
number of physical weight lattice points on the appropriate ``energy shell''.

The eigenfunctions of the Hamiltonian are given by Jack polynomials, on which there
exists an extensive literature. They are of the form
\beq \label{Psiform}
\begin{aligned}
&\langle x|\vec n\rangle_g\ \equiv\ \Psi_{\vec n}^{(g)}(x) 
\= R_{\vec n}^{(g)}(x) \,\De^g 
\qquad\with\quad R_{\vec n}^{(g)}(x) \= P_{\vec n}^{(g)}(x)\,\prod_k x_k^{-p}\ ,\\[4pt]
&\textrm{where}\qquad
x=(x_1,x_2,\ldots,x_N)\ ,\qquad
|\vec n|=n_1{+}n_2{+}\ldots{+}n_{N-1}\ ,\qquad
p=|\vec n|/N\ ,
\end{aligned}
\eeq
and $P_{\vec n}^{(g)}$ is a homogeneous per\-mu\-ta\-tion-symmetric polynomial 
of degree $|\vec n|$ in~$x$. The rational function~$R_{\vec n}^{(g)}$ is 
homogeneous of degree~zero, but in the center-of-mass frame we have
$R_{\vec n}^{(g)}=P_{\vec n}^{(g)}$.
The function $R_{\vec n}^{(g)}$ is a linear combination of symmetric basis functions
\beq
\begin{aligned}
&Q^{+}_{\vec m}(x) \= x_1^{m_1-p} x_2^{m_2-p} \cdots x_N^{m_N-p} 
+ \ \textrm{all permutations} \\[4pt]
&\textrm{with}\qquad p=|\vec m|/N \and |\vec m|=|\vec n|{-}\ell N \for \ell=0,1,\ldots
\end{aligned}
\eeq
(remember we put $m_N=0$).
The coefficients are rational functions of the coupling~$g$.
At $g{=}0$ only the leading term remains, 
and $P_{\vec n}^{(0)}\propto Q^{+}_{\vec n}$.
Hence, all functions are Laurent polynomials in the variables $y_i=x_i^{1/N}$.
The structure will become clear from the examples below.

Before the $\cal PT$ deformation, $\De\propto\prod_{i<j}x_{ij}$ vanishes at coinciding 
coordinate values (the Weyl-alcove walls), which renders the wave functions~(\ref{Psiform}) 
non-square-integrable when $g<0$. Therefore, the physical spectrum is empty there.
However, due to the $g\leftrightarrow1{-}g$ symmetry of the Hamiltonian, we should
consider the two ``mirror values'' of~$g$ together to form a single Hilbert space~${\cal H}_g$.
Then, for a given value of $g>\sfrac12$, a generic deformation (with all $\eps_{ij}$ nonzero) 
will abruptly add a second infinite set of energy eigenstates to the spectrum, given
by replacing $g$ with $1{-}g$. Their energies are given by $E_{\vec n}(1{-}g)$ from
(\ref{E1}) or (\ref{E2}) for a second set of quantum numbers~$\vec n$. This produces
a second ``energy shell'', which may carry states all the way down to zero energy 
(if $\vec\mu(1{-}g)$ is located in the physical Weyl chamber). 
For particular (typical integer) values of~$g$ the two shells may possess simultaneous
states, leading to an enhancement of energy degeneracy.
We shall illustrate these features in the examples below.

\section{Conserved charges and intertwiners}

A key tool in the construction of the spectrum and conserved charges 
is the Dunkl operator~\footnote{
For convenience we restrict to $A$-type models in the section. 
Section~6 deals with a more general case.}
\beq
D_i(g) \= \frac{\pa}{\pa q_i} - g\sum_{j(\neq i)} \cot\sfrac{\pi}{L}(q_i{-}q_j)\,s_{ij}
\= \frac{\im\pi}{L}\Bigl[ 2\,x_i\pa_i - g\sum_{j(\neq i)} \frac{x_i{+}x_j}{x_i{-}x_j}\,s_{ij}\Bigr]\ ,
\eeq
where the reflection $s_{ij}$ acts on its right by permuting labels $i$ and~$j$.
It obeys a simple commutation relation,
\beq
\bigl[D_i(g),D_j(g)\bigr] \= -g^2\sfrac{\pi^2}{L^2} \sum_k (s_{ijk}-s_{jik}) \ ,
\quad\textrm{where}\quad s_{ijk}=s_{ij}s_{jk}
\eeq
effects a cyclic permutation of the labels $i,j,k$.

The importance of the Dunkl operator is twofold.
First, any permutation-invariant (in general: Weyl-invariant) polynomial 
of some degree~$k$ in $\{D_i\}$ will, when restricted to totally symmetric functions, 
give rise to a Liouville charge~$C_k$, i.e.\ a conserved quantity in involution. 
A simple basis of this ring is provided by the Newton sums,
\beq \label{Newtonsums}
I_k(g) \= \textrm{res}\sum_i D_i(g)^k
\qquad\Rightarrow\qquad \bigl[I_k(g),I_\ell(g)\bigr]=0\ ,
\eeq
where `res' denotes the restriction to totally symmetric functions, giving
\beq
I_k(g)\,H(g) \= H(g)\,I_k(g)\ .
\eeq
The total momentum and the Hamiltonian itself are the prime examples,
\beq \label{I1I2}
I_1(g) \= \im P = \sfrac{2\im\pi}{L}\sum_i x_i\pa_i  \und
I_2(g) \= -2\,\bigl(H(g) -E_0(g)\bigr) \ .
\eeq
In the center of mass, $P=0$ of course. 
Only the first $N$ charges are functionally independent; 
any $I_{k>N}$ can be expressed in terms of these.
The $I_k$ for $3\le k\le N$ may be employed to lift the degeneracy 
of the state labelling by energy alone.

Second, the symmetric restriction of any {\sl anti\/}-invariant polynomial 
of some degree~$k$ in the Dunkl operators will yield an {\sl intertwining\/} operator 
(or shift operator)~$M_k(g)$, obeying
\beq \label{MHintertwine}
M_k(g)\,H(g) \= H(-g)\,M_k(g) \= H(g{+}1)\,M_k(g)\ .
\eeq
The simplest such intertwiner is
\beq
M_{\bar{k}}(g) \= 
\textrm{res}\quad\frac{1}{\bar{k}!}\!\!\!\!\!\!\!\!
\sum_{\textrm{permutations}} \prod_{i<j} \bigl(D_i(g)-D_j(g)\bigr)
\quad\with \bar{k}=\sfrac12N(N{-}1)\ ,
\eeq
where the sum is over all permutations of the $\bar k$ factors in the product.
Comparing (\ref{MHintertwine}) with (\ref{E2}) it can be inferred that the action of 
$M_{\bar{k}}(g)$ on the states is
\beq
M_{\bar{k}}(g)\,|\vec n\rangle_g \ \propto\ |\vec n{-}\vec\delta\rangle_{g+1}
\quad\with \vec\delta\=(N{-}1,N{-}2,\ldots,2,1)\ ,
\eeq
which will vanish if the target quantum numbers no longer respect the restriction in~(\ref{norder}).
The shift operator translates the energy shell by the vector~$\vec\mu$. 
Its repeated action will eventually get the state~$|\vec n\rangle_g$ to the edge of the
physical Weyl chamber. Therefore, any state gets mapped to zero after a certain number of shifts.
The adjoint intertwiner $M_k^\dagger(g)=M_k(-g)$ has the opposite action, 
$\vec n\mapsto\vec n{+}\vec\delta$ while $g\mapsto g{-}1$. Since $M_k$ has a nonzero kernel,
$M_k^\dagger$ is not surjective.

The Liouville charges $I_\ell$ together with the intertwiners $M_k$ form a larger algebra,
which is of interest. Beyond the total momentum and the Hamiltonian, 
the higher conserved charges~(\ref{Newtonsums}) do not intertwine homogeneously 
but mix when $M_k$ is passed through them,
\beq
M_k(g)\,I_\ell(g) \= \bigl[ I_\ell(g{+}1) + \sum_{m<\ell} c_{km}(g)\,I_m(g{+}1)\bigr]\,M_k(g)\ ,
\eeq
with some coefficients $c_{km}(g)$ polynomial in~$g$.
However, it may be possible to find another basis $\{C_k(g)\}$ which intertwines nicely,
\beq
M_k(g)\,C_\ell(g) \= C_\ell(g{+}1)\,M_k(g)\ ,
\eeq
meaning that the shift effected by $M_k$ will map simultaneous eigenstates of the 
whole set $\{C_k\}$ to each other.
The composition $M_k^\dagger(g)M_k(g)$ is by construction an element of the Liouville ring
and thus can be expressed in terms of the $I_k(g)$ (or $C_k(g)$). 

When $g\in\NN$, the energy levels (let us call them `even') are degenerate 
with some at coupling $1{-}g\in-\NN_0$ (call those `odd').
In this case, there exists an extra, odd, conserved charge~\footnote{
Each $M$ factor could even carry a different index~$k$ but the resulting $Q$ charges
presumably differ only by Liouville-charge factors.}
\beq \label{Qdef}
Q_k(g) \= M_k(g{-}1) M_k(g{-}2) \cdots M_k(2{-}g) M_k(1{-}g)
\eeq
mapping ${\cal H}_g$ to itself after fusing the spectra at couplings $g$ and $1{-}g$.
We note that the action of~$Q_k$ is well defined only after applying the ${\cal PT}$ deformation,
since the undeformed spectrum is empty for negative $g$~values.
The hidden supersymmetry operator~$Q_k$ maps between `even' and `odd' states 
in the joined spectrum, which arise from the originally positive and negative $g$~values, respectively.
Its square is a polynomial in the (even) Liouville charges, as will be seen in an example
in~(\ref{Qsquared}) below.

\newpage

\section{Details of the $A_2$ model}

\noindent
In this section we work out the details of the simplest nontrivial case, 
which describes three particles on a circle interacting according to the $A_2$ structure.
For simplicity we put $L=\pi$ from now on; the dimensions can easily be reinstated.
The Hamiltonian in the center-of-mass frame then reads (deformation superscript `$\eps$' suppressed)
\beq \label{A2Ham}
H(g) \= 2\sum^3_{i=1} (x_i\pa_i)^2
-4\,g(g{-}1)\sum_{i<j}\frac{x_i\,x_j}{(x_i{-}x_j)^2}
\= -\sfrac12 I_2(g) + 4g^2\ .
\eeq
The other two conserved charges are
\beq
\begin{aligned}
\sfrac{1}{\im}\,I_1(g) &\= 2\sum_i x_i\pa_i \= P \und \\
\sfrac{\im}{8}\,I_3(g) &\=(x_1\pa_1)^3 -
\Bigl( 3 g(g{-}1) \Bigl[\frac{ x_1\,x_2}{(x_1{-}x_2)^2}+\frac{x_1\,x_3}{(x_1{-}x_3)^2}\Bigr] +
2 g^2\Bigr)\,x_1\pa_1 +\,\textrm{cyclic}\ ,
\end{aligned}
\eeq
but $I_4$ is already dependent,
\beq
I_4\=\sfrac43 I_3 I_1+\sfrac12 I_2^2-I_2 I_1^2+\sfrac16 I_1^4-g^2 I_2 +\sfrac13 g^3 I_1^2\ .
\eeq

The energy formul\ae\ (\ref{E1}) and (\ref{E2}) specialize to
\beq \label{A2energies}
\begin{aligned}
E_{n_1,n_2}(g) &\= 
\sfrac43(n_1^2+n_2^2-n_1n_2) + 4g\,n_1 + 4 g^2\\[4pt]
&\= (n_1+2g)^2 + \sfrac13(n_1-2n_2)^2
\quad\with n_1\ge n_2 \ge 0\ ,
\end{aligned}
\eeq
and the ground state for $g\ge0$ is
\beq
\langle x|0,0\rangle_g \ \equiv\ \Psi_0^{(g)}(x) \= \De^g
\= \Bigl(\frac{x_{12}x_{13}x_{23}}{x_1\,x_2\,x_3}\Bigr)^g 
\= (Q^{-}_{2,1})^g
\quad\with E_0(g) = 4\,g^2\ .
\eeq
Here, we introduced (for later purposes) the antisymmetric basis functions, so
\beq
Q^{\pm}_{m_1,m_2}\= x_1^{m_1-p}x_2^{m_2-p}x_3^{-p}\pm x_1^{m_2-p}x_2^{m_1-p}x_3^{-p}
+ \ \textrm{cyclic} \quad\with p=(m_1{+}m_2)/3\ .
\eeq
These Laurent polynomials (in $x_i^{1/3}$) form a ring whose structure we detail in Appendix~B.
In Appendix~C we list the explicit wave functions
\beq
\Psi^{(g)}_{n_1,n_2}(x) \= R^{(g)}_{n_1,n_2}(x)\,\Psi_0^{(g)}(x)
\eeq
(see (\ref{Psiform})) for small values of $n_1$.

\begin{figure}[h!]
\centering
\includegraphics[scale=0.8]{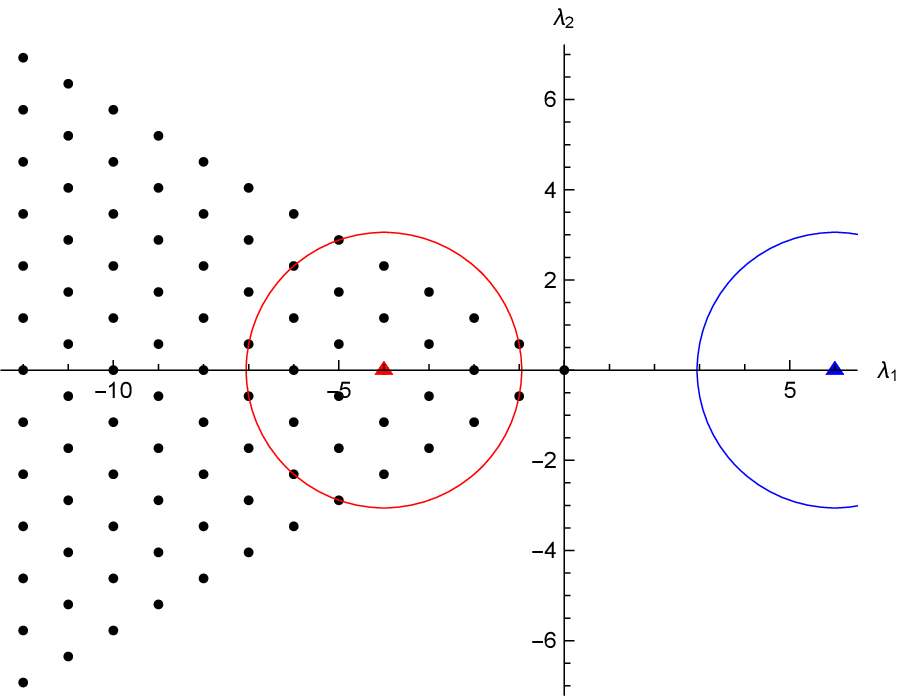}
\qquad
\includegraphics[scale=0.8]{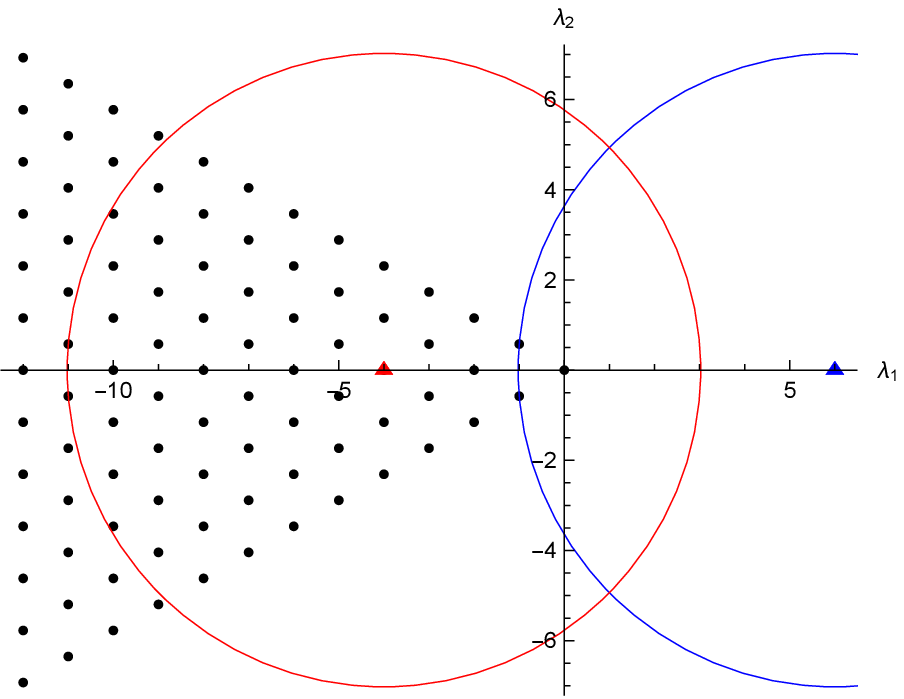}
\caption{States (black dots) and ``energy shells'' (blue for $g{=}3$, red for $g{=}{-}2$) 
in weight space for $E=\sfrac{28}{3}$ (left) and $E=\sfrac{148}{3}$ (right).}
\label{fig1}
\end{figure}

\begin{figure}[h!]
\centering
\includegraphics[scale=0.6]{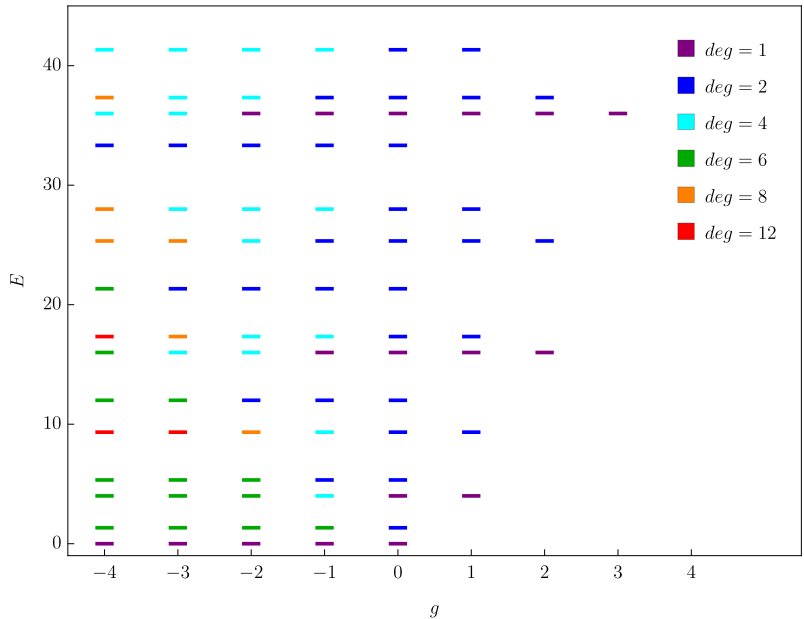}
\caption{Low-lying energy spectrum and degeneracies (color-coded) for the $A_2$ model at integer coupling $g\in\Z$
in the range $g\in[-4,+4]$.}
\label{fig2}
\end{figure}

Each eigenstate $|n_1,n_2\rangle$ corresponds to a point
\beq
(\lambda_1,\lambda_2) = \bigl(-n_1,\sfrac{1}{\sqrt{3}}(n_1{-}2n_2)\bigr)
\quad\with \lambda_1\le-\sqrt{3}\,|\lambda_2|
\eeq
in a $\sfrac{\pi}{3}$ wedge around the negative $\lambda_1$ axis.
The circles determining the energy eigenstates for couplings $g$ and $1{-}g$
are centered at 
\beq
(2,0)\,g \und (2,0)\,(1{-}g)
\eeq
in $\lambda$-space, respectively. This is illustrated in Figure~\ref{fig1}.
Since $\mu_2=0$, we have an obvious energy degeneracy for 
\beq
|n_1,n_2\rangle_g \and |n_1,n_1{-}n_2\rangle_g
\eeq
except for $n_1{=}2n_2$, of course. 
For $g\ge0$ there rarely appears higher degeneracy,\footnote{
Occasional energy values are triply or quadruply degenerate.}
but at $g<0$ energy levels
are up to 12-fold degenerate! 
This plethora of states becomes physical only after the $\cal PT$ deformation
and greatly enlarges the Hilbert space~${\cal H}_g$ for any $g{>}0$.
Figure~\ref{fig2} displays the energy spectra with degeneracies
for low levels and small integral values of~$g$.

The basic intertwiner for the $A_2$ model is of order three,
\beq
M_3(g) \= \sfrac13\,\textrm{res}\,\bigl( 
D_{12}D_{23}D_{31}+D_{23}D_{31}D_{12}+D_{31}D_{12}D_{23} \bigr)(g)
\eeq
with the obvious notation $D_{ij}=D_i{-}D_j$. It computes to
\beq
\begin{aligned}
M_3(g) &\= \pa_{12}\pa_{23}\pa_{31} - 
2g\,\bigl\{\cot q_{12}\pa_{23}\pa_{31}+\textrm{cyclic}\bigr\} +
4g^2\bigl(\{\cot q_{12}\cot q_{23}\pa_{31}+\textrm{cyclic}\bigr\}
\\[2pt] &\quad-
g(g{-}1)\bigl\{\sin^{-2}\!q_{31}\pa_{31}+\textrm{cyclic}\bigr\} -
4g(2g{-}1)\bigl\{\sin q_{12}\sin q_{23}\sin q_{31}\bigr\}^{-1} 
\\[4pt] &\quad-
2g(3g^2{-}g{+}2) \cot q_{12}\cot q_{23}\cot q_{31} +
2g(g{-}1)(g{+}2) \bigl\{\cot^3 q_{12}+\textrm{cyclic}\bigr\}
\end{aligned}
\eeq
where we abbreviated $\pa_{ij}=\sfrac{\pa}{\pa q_i}{-}\sfrac{\pa}{\pa q_j}$
and $q_{ij}=q_i{-}q_j$.
In terms of the multiplicative variables, the shift operator takes the form
\beq
\begin{aligned}
M_3(g)
&\ \propto\ (x_1\pa_1{-}x_2\pa_2)(x_2\pa_2{-}x_3\pa_3)(x_3\pa_3{-}x_1\pa_1)-
g\bigl\{\sfrac{x_1+x_2}{x_1-x_2}(x_2\pa_2{-}x_3\pa_3)(x_3\pa_3{-}x_1\pa_1)
+\textrm{cyclic}\bigr\} \\[6pt] &\quad +
g^2\bigl\{\sfrac{x_1+x_2}{x_1-x_2}\sfrac{x_2+x_3}{x_2-x_3}(x_3\pa_3{-}x_1\pa_1)
+\textrm{cyclic}\bigr\} -
g(g{-}1)\bigl\{\sfrac{x_3 x_1}{x_{31}^2}(x_3\pa_3{-}x_1\pa_1)
+\textrm{cyclic}\bigr\} \\[4pt] &\quad +
4g(2g{-}1)\sfrac{x_1 x_2 x_3}{x_{12}x_{23}x_{31}} -
\sfrac14g(3g^2{-}g{+}2)
\sfrac{x_1+x_2}{x_1-x_2}\sfrac{x_2+x_3}{x_2-x_3}\sfrac{x_3+x_1}{x_3-x_1} 
\\[4pt] &\quad +
\sfrac14g(g{-}1)(g{+}2)\bigl\{\bigl(\sfrac{x_1{+}x_2}{x_1{-}x_2}\bigr)^3
+\textrm{cyclic}\bigr\}\ .
\end{aligned}
\eeq
It action on the states is
\beq \label{stateshift}
M_3(g)\,|n_1,n_2\rangle_g \ \propto\ |n_1{-}2,n_2{-}1\rangle_{g+1}
\eeq
conserving the energy. In weight space it moves 
$(\lambda_1,\lambda_2)\mapsto(\lambda_1{+}2,\lambda_2)$.

For homogeneous intertwining relations, we redefine
\beq
C_1 = I_1\ ,\quad C_2 = I_2 -2E_0 = -2\,H \and C_3 = I_3 - I_1 I_2
\eeq
which obey
\beq
M_3(g)\,C_k(g) \= C_k(g{+}1)\,M_3(g)\ .
\eeq
This imples that the eigenvalue of~$C_3$ is also conserved under the shift action.
Indeed, it is readily verified that
\beq 
\begin{aligned}
C_1\,|n_1,n_2\rangle_g \= 0 \ ,\qquad
&C_2(g)\,|n_1,n_2\rangle_g \= -2\bigl[(n_1{+}2g)^2+\sfrac13(n_1{-}2n_2)^2\bigr]\,|n_1,n_2\rangle_g\ ,\\[4pt]
C_3(g)\,|n_1,n_2\rangle_g &\= 
-\sfrac89\im\,(n_1{-}2n_2)(2n_1{-}n_2{+}3g)(n_1{+}n_2{+}3g)\,|n_1,n_2\rangle_g\ ,
\end{aligned}
\eeq
which is compatible with the shift~(\ref{stateshift}).
The degeneracy reflection $n_2\mapsto n_1{-}n_2$ flips the sign of~$C_3$,
so two such states can be discriminated by their $C_3$ eigenvalues.

As expected, the composition of the intertwiner with its adjoint yields an expression
in the Liouville charges,
\beq
\begin{aligned}
M_3^\dagger M_3(g) &\ \propto\
18I_3^2{-}36I_3I_2I_1{+}8I_3I_1^3{-}3I_2^3{+}21I_2^2I_1^2{-}9I_2I_1^4{+}I_1^6
+2g^2(9I_2^2{-}6I_2I_1^2{+}I_1^4) \\[4pt]
&\= 18C_3^2+8C_3C_1^3-3C_2^3+3C_2^2C_1^2-C_2C_1^4+C_1^6\\
&\qquad -6g^2(3C_2-C_1^2+8g^2)^2 \ .
\end{aligned}
\eeq

Let us take a look at the extra degeneracy between even ($g{>}0$) and odd ($g{\le}0$) states
appearing when $g\in\Z$. The odd operator~$Q_3(g)$ mapping one to the other and defined 
in~(\ref{Qdef}) is of order $3(2g{-}1)$ and shifts the quantum numbers as
\beq
\bigl(n_1,n_2\bigr)\ \mapsto\ \bigl(n_1{-}4g{+}2,n_2{-}2g{+}1\bigr)\ ,
\eeq
which produces a rather large kernel.
$Q_3$ commutes with all conserved charges~$I_k$, so it keeps their eigenvalues.
In weight space, it maps between the `even' and `odd' energy shells.

Finally, we briefly discuss two other kinds of $\cal PT$ deformations in the $A_2$-model context,
which we denote as `angular' and `radial', respectively.
Different from the parity transformation in~(\ref{PT1}), which amounts to the outer conjugation
automorphism of $A_{N-1}$, the angular and radial deformations are compatible with an
elementary Coxeter reflection (or particle permutation), e.g.
\beq
{\cal P} : \quad (q_1,q_2,q_3)\mapsto(q_2,q_1,q_3) \und
(x_1,x_2,x_3)\mapsto(x_2,x_1,x_3)\ ,
\eeq
while $\cal T$ remains complex conjugation.

The angular $\cal PT$ deformation is homogeneous in the $q_i$ coordinates,
in contrast to the constant complex coordinate shift~(\ref{PTshift}).
It is induced by a complex orthogonal coordinate change,
$\vec q\mapsto \Gamma_\eps\vec q$ with $\Gamma_\eps\in\textrm{SO}(3,\C)$ modulo 
SO$(3,\R)$, described in~\cite{CoLe17}. Explicitly,
\beq
q_i\ \mapsto\ q_i^\eps \= \sfrac13\bigl[ (1{+}2\cosh\eps)\,q_i + 
(1{-}\cosh\eps-\im\sqrt{3}\sinh\eps)\,q_j + (1{-}\cosh\eps+\im\sqrt{3}\sinh\eps)\,q_k \bigr]
\eeq
with $(i,j,k)$ being a cyclic permutation of~$(1,2,3)$. 
This deformation does not entirely remove the singular loci of the potential given by
\beq
\begin{aligned}
0\=\sin q_{ij}^\eps &\=
\cosh(\sqrt{3}\sinh\eps\,q_k)\,\sin(\cosh\eps\,q_{ij} )+ \im\,
\sinh(\sqrt{3}\sinh\eps\,q_k)\,\cos(\cosh\eps\,q_{ij}) \\[4pt]
&\ \Leftrightarrow\qquad q_{ij} = \frac{\ell\,\pi}{\cosh\eps} 
\quad\wedge\quad q_k = 0
\quad\for \ell=0,1,2,\ldots\ ,
\end{aligned}
\eeq
where again $(i,j,k)$ are cyclic and we went to the conter-of-mass frame,
so $q_{ik}+q_{jk}=-3q_k$.
For small enough~$\eps$, only the origin $\{q_i=0\}$ remains singular,
but with growing value of~$\eps$ extra singularities appear inside the Weyl alcove. 

The radial $\cal PT$ deformation is a nonlinear one,
\beq
q_i\ \mapsto\ q_i^\eps \= q_i + \im\eps\,q_{jk}/r
\quad\with r^2=q_{12}^2+q_{23}^2+q_{31}^2
\eeq
and $(i,j,k)$ being cyclic once more. 
The remaining singularities occur ar
\beq
q_{ij} = \ell\,\pi \quad\wedge\quad q_k = 0 \quad\for \ell=0,1,2,\ldots\ ,
\eeq
and in addition one should average the potential,
$V\mapsto V_\eps+V_{-\eps}$ with $V_\eps(q)=V(q^\eps)$. 
Both cases can be parametrized jointly by writing
\beq
V_\eps(q_{ij}) \= V\bigl( R(\eps)\,q_{ij} + \im I(\eps)\,q_k\bigr)
\quad\with\begin{cases}
\ph R(\eps)=\cosh\eps \!\!\!\!\!\!&\and I(\eps)=\sqrt{3}\sinh\eps \\[4pt]
\ph R(\eps)\=1 &\and I(\eps)=-3/r
\end{cases}
\eeq
for the angular and radial $\cal PT$ deformation, respectively.

\section{Details of the $G_2$ model}

\noindent
For a more complicated and non-simply-laced example, 
we turn to the $G_2$ model~\cite{Qu95,Qu96} for three particles on a circle 
and apply the constant-shift $\cal PT$ deformation~(\ref{PTshift})
but suppress it notationally.
The $G_2$ model adds to the previous two-body potential 
of the $A_2$ case~(\ref{A2Ham}) a specific three-body interaction,
\beq \label{G2Ham}
\begin{aligned}
H(g)&\=-\frac{1}{2}\sum^3_{i=1} \frac{\pa}{\pa q_i^2}
+\sum_{i<j}\frac{\gs(\gs{-}1)}{\sin^2(q_i{-}q_j)} 
+\sum_{i<j}\frac{3\gl(\gl{-}1)}{\sin^2(q_i{+}q_j{-}2q_k)} \\[4pt]
&\=2\sum^3_{i=1} (x_i\pa_i)^2
-4\,\gs(\gs{-}1)\sum_{i<j}\frac{x_i\,x_j}{(x_i{-}x_j)^2}
-12\,\gl(\gl{-}1)\sum_{i<j}\frac{x_i\,x_j\,x_k^2}{(x_i x_j{-}x_k^2)^2}\ ,
\end{aligned}
\eeq
where the index `$k$' complements $i$ and $j$ to the triple (1,2,3),
there are two independent real couplings $\gs$ and~$\gl$,
and we again put $L=\pi$ for simplicity.
The potential can be viewed as a sum of two copies of the $A_2$ potential,
with a relative coordinate rotation between them. 
The singular walls appear for
\beq
q_i-q_j=0 \und q_k=\sfrac13(q_1{+}q_2{+}q_3) \quad\for i,j,k\in\{1,2,3\}\ ,
\eeq
bounding the $G_2$ Weyl chambers.
The Weyl group is enhanced from $S_3$ to $D_6$, generated by
\beq \label{G2Coxq}
s_{12} : \ (q_1,q_2,q_3) \mapsto (q_2,q_1,q_3) \and
\jmath : \ (q_1,q_2,q_3) \mapsto 
\sfrac23(q_1{+}q_2{+}q_3)(1,1,1) - (q_1,q_2,q_3)
\eeq
and permutations, which for the $x_i$ coordinates translates to
\beq \label{G2Coxx}
s_{12} : \ (x_1,x_2,x_3) \mapsto (x_2,x_1,x_3) \and
\jmath : \ (x_1,x_2,x_3) \mapsto (x_1x_2x_3)^{2/3}(\sfrac1{x_1},\sfrac1{x_2},\sfrac1{x_3})\ .
\eeq 

The Hamiltonian (\ref{G2Ham}) yields eigenvalues
\beq
\begin{aligned} \label{G2energies}
E_{n_1,n_2}(g) &\= \sfrac43(n_1^2+n_2^2-n_1n_2) + 
4\gs\,n_1 + 4\gl\,(2n_1-n_2)+ 4(\gss+3\gls+3\gs\gl)\\[4pt]
&\= (n_1+2\gs+3\gl)^2 + \sfrac13(n_1-2n_2+3\gl)^2
\quad\with n_1 \ge 2n_2 \ge 0\ ,
\end{aligned}
\eeq
and the ground-state wave function for $\gs\ge0$ and $\gl\ge0$ is 
\beq \label{G2ground}
\langle x|0,0\rangle_{\gs,\gl} \ \equiv\ \Psi_0^{(\gs,\gl)}(x) \= 
\De_{\textrm{S}}^{\gs} \De_{\textrm{L}}^{\gl}
\quad\with E_0(g) = 4\,(\gss+3\gls+3\gs\gl)\ ,
\eeq
where we introduced
\beq \label{G2Deltas}
\begin{aligned}
\De_{\textrm{S}} &\= \frac{(x_1{-}x_2)(x_1{-}x_3)(x_2{-}x_3)}{x_1\,x_2\,x_3} \= Q^-_{2,1}\ ,\\[4pt]
\De_{\textrm{L}} &\= \frac{(x_1^2{-}x_2x_3)(x_2^2{-}x_1x_3)(x_3^2{-}x_1x_2)}{x_1^2\,x_2^2\,x_3^2}
\= \sfrac12\bigl( Q^+_{3,0}-Q^+_{3,3}\bigr)\ .
\end{aligned}
\eeq
In addition to the permutation symmetry inherited from the $A_2$ model,
we also have to impose (anti-)invariance under the additional (even) Coxeter 
element~$\jmath$ in~(\ref{G2Coxx}), which implements an inversion in $x$~space
and flips the sign of the roots by a $\pi$ rotation in the relative $q$~space.
Noting that $\jmath^2=1$ and $[\jmath,s_{ij}]=0$ and
\beq \label{Deltashifts}
j : Q^{\pm}_{m_1,m_2}\ \mapsto\ \pm Q^{\pm}_{m_1,m_1{-}m_2}
\eeq
one sees that $j$ flips the sign of both $\De_S$ and $\De_L$, hence
\beq
j\,\Psi_0^{(\gs,\gl)}(x) \= (-)^{\gs+\gl} \Psi_0^{(\gs,\gl)}(x)\ .
\eeq
We also deduce that the energy-degenerate states 
$|n_1,n_2\rangle$ and $|n_1,n_1{-}n_2\rangle$
of the $A_2$ model are related by the action of~$j$.
Therefore, only their sum or difference will be a $G_2$-model state,
so the range of $n_2$ can be restricted to $n_2\le\sfrac12n_1$, 
as already claimed in~(\ref{G2energies}).

The excited states then take the form
\beq \label{G2states}
\Psi_{n_1,n_2}^{(\gs,\gl)}(x) \=
R_{n_1,n_2}^{(\gs,\gl)}(x)\,\Psi_0^{(\gs,\gl)}(x)\ ,
\eeq
where the $R_{n_1,n_2}^{(\gs,\gl)}$ are again particular Weyl-symmetric rational functions 
of degree zero. Appendix~D contains a list of low-lying wave functions.

\begin{figure}[h!]
\centering
\includegraphics[scale=0.8]{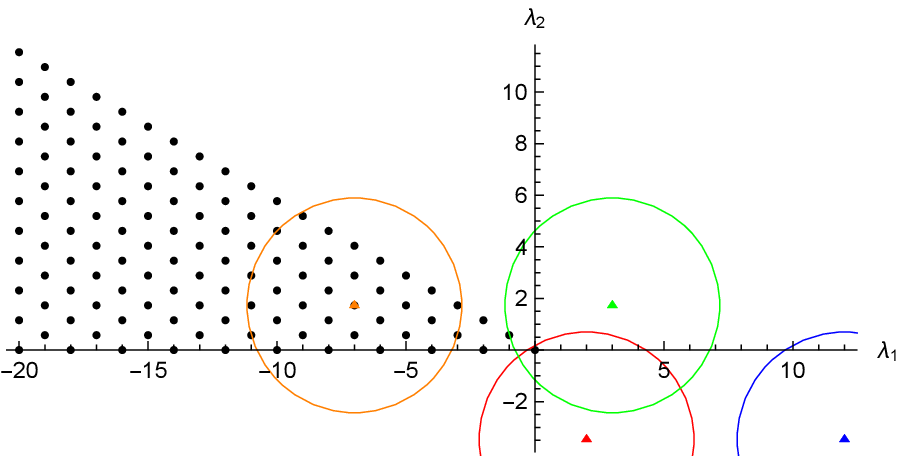}
\qquad
\includegraphics[scale=0.8]{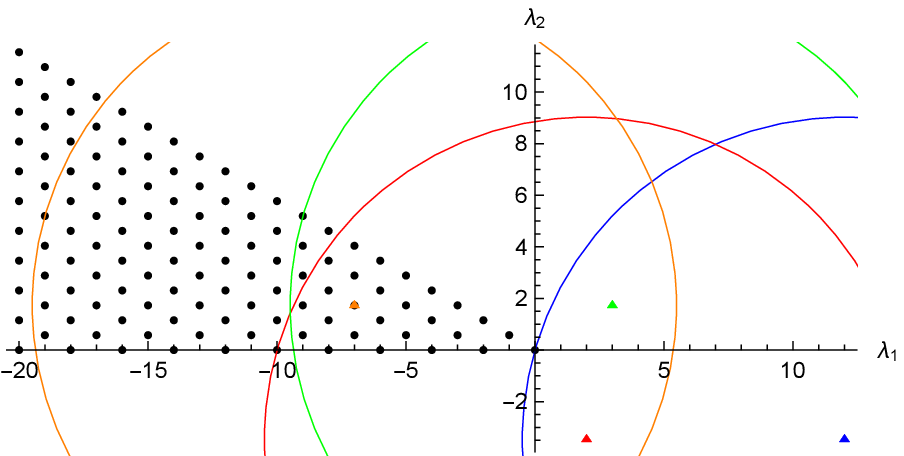}
\caption{States (black dots) and ``energy shells'' 
for $(\gs,\gl)\,{=}(3,2)$ (blue), ${=}({-}2,2)$ (red), 
${=}(3,{-}1)$ (green) and ${=}({-}2,{-}1)$ (orange)
for $E=\sfrac{52}{3}$ (left) and $E=156$ (right).}
\label{fig3}
\end{figure}

Each eigenstate $|n_1,n_2\rangle$ corresponds to a point
\beq
(\lambda_1,\lambda_2) = \bigl(-n_1,\sfrac{1}{\sqrt{3}}(n_1{-}2n_2)\bigr)
\quad\with \lambda_1\le-\sqrt{3}\,\lambda_2\le 0
\eeq
in a $\sfrac{\pi}{6}$ wedge above the negative $\lambda_1$ axis,
in accord with one $G_2$ Weyl chamber.
The circles determining the energy eigenstates for couplings 
$(\gs,\gl)$, $(1{-}\gs,\gl)$, $(\gs,1{-}\gl)$ and $(1{-}\gs,1{-}\gl)$
are centered at 
\beq
\begin{aligned}
(2\gs{+}3\gl,-\sqrt{3}\gl)\ ,\quad
(2{-}2\gs{+}3\gl,-\sqrt{3}\gl)\ ,\quad \\
(3{+}2\gs{-}3\gl,\sqrt{3}\gl{-}\sqrt{3})\ ,\quad
(5{-}2\gs{-}3\gl,\sqrt{3}\gl{-}\sqrt{3})
\end{aligned}
\eeq
in $\lambda$-space, respectively. This is illustrated in Figure~\ref{fig3}.
After the $\cal PT$ deformation, the Hilbert space ${\cal H}_{\gs,\gl}$ comprises the
four towers obtained from the four circles in~Figure~3.
Again, for integral values of the couplings, the towers have matching energy levels,
which greatly increases their degeneracy.

The $G_2$ Dunkl operator is an extension of the $A_2$ one 
(again with $\{i,j,k\}=\{1,2,3\}$),
\beq
-\im\,D_i(g) \= 
2\,x_i\pa_i - \gs\sum_{j(\neq i)} \frac{x_i{+}x_j}{x_i{-}x_j}\,s_{ij} -
\gl\biggl( \sum_{j(\neq i)}  \frac{x_i x_j{+}x_k^2}{x_i x_j{-}x_k^2}\,s_{ik} -
2\,\frac{x_j x_k{+}x_i^2}{x_j x_k{-}x_i^2}\,s_{jk}\biggr)\jmath\ .
\eeq
The first two Newton sums in this Dunkl operator yield the conserved 
momentum and energy,
\beq
I_1 \= \textrm{res}\sum_i D_i \= \im\,P \und
I_2 \= \textrm{res}\sum_i D_i^2 \= -2(H-E_0)\ ,
\eeq
because $\sum_i q_i$ and $\sum_i q_i^2$ are not only permutation-symmetric
but also invariant under the rotation~$\jmath$ from~(\ref{G2Coxq}).
This, however, is not the case for $\sum_i q_i^r$ when $r{\ge}3$,
but one can find a Weyl-invariant combination at order six,
\beq
\si_3\,\jmath\,\si_3 \= \si_3 (\si_3 - 2\si_2\si_1 + \sfrac49\si_1^3 )
\quad\with \si_r =\textstyle{\sum_i} q_i^r\ ,
\eeq
which generates another Liouville charge,
\beq
\begin{aligned}
&J_6 \= \textrm{res}\bigl\{ 
5D_1^6+12D_1^5D_2+12D_1^4D_2^2-24D_1^4D_2D_3+10D_1^3D_2^3+12D_1^3D_2^2D_3 \\[4pt] 
&\qquad\qquad +\ \textrm{permutations(1,2,3)} \bigr\}_\textrm{symmetrized} \\[4pt]
&\=5 (x_1\pa_1)^6+12 (x_1\pa_1)^5 x_2\pa_2+12 (x_1\pa_1)^4 (x_2\pa_2)^2\\
&\quad -24 (x_1\pa_1)^4 (x_2\pa_2)(x_3\pa_3)+10 (x_1\pa_1)^3 (x_2\pa_2)^3 +12 (x_1\pa_1)^3 (x_2\pa_2)^2(x_3\pa_3)  \\ 
&\quad +c_{400}{\cdot}(x_1\pa_1)^4 + c_{310}{\cdot}(x_1\pa_1)^3 x_2\pa_2 +c_{220}{\cdot}(x_1\pa_1)^2(x_2\pa_2)^2 +c_{211}{\cdot}(x_1\pa_1)^2 (x_2\pa_2 )(x_3\pa_3) \\ 
&\quad +c_{300}{\cdot}(x_1\pa_1)^3+ c_{210}{\cdot}(x_1\pa_1)^2(x_2\pa_2)+c_{111}{\cdot}(x_1\pa_1) (x_2\pa_2 )(x_3\pa_3) \\ 
&\quad +c_{200}{\cdot}(x_1\pa_1)^2+ c_{110}{\cdot}(x_1\pa_1)(x_2\pa_2)+ c_{100}{\cdot}(x_1\pa_1)+c_{000}+\ \textrm{permutations(1,2,3)}
\end{aligned}\label{Jsix}
\eeq
where symmetrization means Weyl ordering of every summand, 
and the coefficients $c_{s_1s_2s_3}(x)$ are given in Appendix~E.

From~(\ref{Deltashifts}) we see that 
the $G_2$ model enjoys two separate intertwiners,
\beq
\begin{aligned}
M_{3,\textrm{S}} &\= \sfrac13\,\textrm{res}\,
(D_1{-}D_2)(D_2{-}D_3)(D_3{-}D_1)\ + \textrm{cyclic}\ ,\\[4pt]
M_{3,\textrm{L}} &\= \sfrac13\,\textrm{res}\,
(D_1{+}D_2{-}2D_3)(D_2{+}D_3{-}2D_1)(D_3{+}D_1{-}2D_2)
\ + \textrm{cyclic}\ ,
\end{aligned}
\eeq
which independently shift by unity the couplings $\gs$ and $\gl$,
respectively,
\beq
\begin{aligned}
M_{3,\textrm{S}} \,|n_1,n_2\rangle_{\gs,\gl} \ &\propto\ 
|n_1{-}2,n_2{-}1\rangle_{\gs+1,\gl} \ ,\\[4pt]
M_{3,\textrm{L}} \,|n_1,n_2\rangle_{\gs,\gl} \ &\propto\ 
|n_1{-}3,n_2\rangle_{\gs,\gl+1} \ .
\end{aligned}
\eeq
Their explicit form is
\beq
\begin{aligned}
M_{3,\textrm{S}} 
&\ \propto\ (x_1\pa_1{-}x_2\pa_2)(x_2\pa_2{-}x_3\pa_3)(x_3\pa_3{-}x_1\pa_1)-
\gs \bigl\{\sfrac{x_1+x_2}{x_1-x_2}(x_2\pa_2{-}x_3\pa_3)(x_3\pa_3{-}x_1\pa_1)
+\textrm{cyclic}\bigr\} \\[6pt] 
&\quad + \gss\bigl\{\sfrac{x_1+x_2}{x_1-x_2}\sfrac{x_2{+}x_3}{x_2{-}x_3}(x_3\pa_3{-}x_1\pa_1)
+\textrm{cyclic}\bigr\} - \gs(\gs{-}1)\bigl\{\sfrac{x_3 x_1}{x_{31}^2}(x_3\pa_3{-}x_1\pa_1)
+\textrm{cyclic}\bigr\} \\[4pt] 
&\quad +9\gl(\gl{-}1)\bigl\{\sfrac{x_2^2 x_3 x_1}{(x_2^2{-}x_3 x_1)^2}(x_3\pa_3{-}x_1\pa_1)
+\textrm{cyclic}\bigr\}+
2\gs(4\gs{-}2 {+}9\gl(\gl{+}1))\sfrac{x_1 x_2 x_3}{x_{12}x_{23}x_{31}} \\[4pt] 
&\quad -\sfrac14\gs(3\gss{-}\gs{+}2) 
\sfrac{x_1{+}x_2}{x_1{-}x_2}\sfrac{x_2+x_3}{x_2-x_3}\sfrac{x_3{+}x_1}{x_3{-}x_1} +
\sfrac14\gs(\gs{-}1)(\gs{+}2)\bigl\{\bigl(\sfrac{x_1{+}x_2}{x_1{-}x_2}\bigr)^3
+\textrm{cyclic}\bigr\} \\[4pt] 
&\quad -9 \gs\gl(\gl{+}1)\bigl\{\sfrac{x_3{+}x_1}{x_3-x_1} \sfrac{x_2^2 x_3 x_1}{(x_2^2{-}x_3 x_1)^2} 
+\textrm{cyclic}\bigr\}\  ,
\end{aligned}
\eeq
\beq
\begin{aligned}
M_{3,\textrm{L}} 
&\ \propto\ (x_1\pa_1{+}x_2\pa_2{-}2x_3 \pa_3)(x_2\pa_2{+}x_3\pa_3{-}2x_1 \pa_1)(x_3\pa_3{+}x_1\pa_1{-}2x_2 \pa_2) \\[6pt] 
&\quad +3 \gl\bigl\{ \sfrac{x_2^2{+}x_3x_1}{x_2^2{-}x_3x_1}(x_1\pa_1{+}x_2\pa_2{-}2x_3 \pa_3)(x_2\pa_2{+}x_3\pa_3{-}2x_1 \pa_1)+\textrm{cyclic}\bigr\} \\[4pt] 
&\quad +9 \gl\bigl\{x_1x_2 \bigl[ \sfrac{ (\gl-1) x_2 x_3}{(x_2^2{-}x_1 x_3)^2}+\sfrac{(\gl-1) x_1 x_3}{(x_1^2{-}x_2 x_3)^2}
+\sfrac{2 \gl(x_3^2{+}x_1 x_2)}{(x_2^2{-}x_1 x_3)(x_1^2{-}x_2 x_3)} \bigr] (x_1\pa_1{+}x_2\pa_2{-}2x_3 \pa_3) +\textrm{cyclic}\bigr\} \\[4pt] 
&\quad +9 \gs(\gs{-}1)\bigl\{ \sfrac{x_1x_2}{(x_1{-}x_2)^2}  (x_1\pa_1{+}x_2\pa_2{-}2x_3 \pa_3) +\textrm{cyclic} \bigr\} 
- 27 \gl\gs(\gs{-}1)\bigl\{ \sfrac{x_1x_2}{(x_1{-}x_2)^2}  +\textrm{cyclic} \bigr\}  \\[4pt] 
&\quad +54\gl \bigl[ \gs(\gs{-}1) \sfrac{(x_1^2{-}x_2 x_3{+}x_2^2{-}x_3 x_1{+}x_3^2{-}x_1 x_2)^3}{(x_1-x_2)^2(x_2-x_3)^2(x_3-x_1)^2}
- \gl(\gl{-}3)\bigr] \sfrac{x_1^2 x_2^2 x_3^2}{(x_1^2{-}x_2 x_3)(x_2^2{-}x_3 x_1)(x_3^2{-}x_1 x_2)} \\[4pt]
&\quad -27 \gl(\gls{+}\gl{-}2) x_1 x_2 x_3 \bigl\{x_1 \sfrac{x_1^2+x_2 x_3}{x_1^2{-}x_2 x_3}+\textrm{cyclic}\bigr\} 
+ 27 \glc \sfrac{(x_1^2{+}x_2 x_3)(x_2^2{+}x_3 x_1)(x_3^2{+}x_1 x_2)}{(x_1^2{-}x_2 x_3)(x_2^2{-}x_3 x_1)(x_3^2{-}x_1 x_2)}\ .
\end{aligned}
\eeq

A better basis for the Liouville charges is
\beq
C_1 = I_1\ ,\quad C_2 = I_2-2E_0 = -2\,H \und
\eeq
\beq
\begin{aligned}%\notag
C_6 &\= J_6 +2 (27 \gls + 24 \gl\gs + 8 \gss)  C_2 C_1^2-9 \gls C_2^2
-\sfrac19 (105 \gls + 96 \gl\gs + 32 \gss) C_1^4 \\[4pt]
&\qquad +16 (39 \glq + 72 \glc\gs + 60 \gls\gss + 24 \gl\gs^3+ 4 \gsq) C_1^2 -144 \glq C_2- 576 \gl^6 \ ,
%&\=J_6 - \sfrac19 (105 \gls + 96 \gl\gs + 32 \gss) (C_1^2-\sfrac92 C_2) C_1^2 + \sfrac32\gls C_2 C_1^2 -9 \gls(C_2+8\gls)^2 \\[4pt]
%&\qquad +16 (39 \glq + 72 \glc\gs + 60 \gls\gss + 24 \gl\gsc + 4 \gsq) C_1^2 \ ,
\end{aligned}
\eeq
obeying homogeneous intertwining relations
\beq
\begin{aligned}
M_{3,\textrm{S}}(\gs,\gl)\,C_k(\gs,\gl) &\= C_k(\gs{+}1,\gl)\,M_{3,\textrm{S}} \ ,\\[4pt]
M_{3,\textrm{L}}(\gs,\gl)\,C_k(\gs,\gl) &\= C_k(\gs,\gl{+}1)\,M_{3,\textrm{L}} \ .
\end{aligned}
\eeq
This is also signified by the action
\beq
C_6\,|n_1,\!n_2\rangle_{\gs,\gl} \=- \sfrac{64}{81} 
(3\gl{+}n_1{-}2 n_2)^2 (6 \gl{+}3 \gs{+}2 n_1{-}n_2)^2 (3\gl{+}3\gs+n_1{+}n_2)^2 |n_1,\!n_2\rangle_{\gs,\gl} .
\eeq
The intertwining with their corresponding conjugates produces two polynomials in the Liouville charges,
\beq
\begin{aligned} \label{polym}
M_{3,\textrm{S}}^\dagger M_{3,\textrm{S}}
&\= -3 C_6 -\sfrac16 C_1^6 +\sfrac32 C_2 C_1^4-\sfrac72 C_2^2 C_1^2 +\sfrac12 C_2^3+\gss(C_1^2-3C_2-8\gss)^2 \ ,\\[4pt]
M_{3,\textrm{L}}^\dagger M_{3,\textrm{L}}
&\= 81 C_6 + C_1^2(2C_1^2-9C_2)^2 + 81\gls(C_1^2-3C_2-24\gls)^2\ . 
\end{aligned}
\eeq
The intertwining operators also enable odd conserved charges when the couplings take integer values, in the form
of the chain of operators
\beq
\begin{aligned}
Q_{3,\textrm{S}}(\gs,\gl) &\= 
M_{3,\textrm{S}}(\gs{-}1,\gl) M_{3,S} (\gs{-}2,\gl) \cdots M_{3,S} (2{-}\gs,\gl) M_{3,S} (1{-}\gs,\gl) \ , \\[4pt]
Q_{3,\textrm{L}}(\gs,\gl) &\= 
M_{3,\textrm{L}}(\gs,\gl{-}1) M_{3,L} (\gs,\gl{-}2) \cdots M_{3,L} (\gs,2{-}\gl) M_{3,L} (\gs,1{-}\gl) \ ,
\end{aligned}
\eeq
which in the simplest non-trivial cases squares to the form of the polynomials in (\ref{polym}), 
\beq \label{Qsquared}
\begin{aligned}
Q_{3,\textrm{S}}^2(2,\gl) &\= 
\left( 3 C_6 +\sfrac16 C_1^6 -\sfrac32 C_2 C_1^4 +\sfrac72 C_2^2 C_1^2 -\sfrac12 C_2^3 \right)^3 + \textrm{lower terms}\ ,\\[4pt]
Q_{3,\textrm{L}}^2(\gs,2) &\= 
\left( 81 C_6 +4C_1^2(C_1^2-\sfrac92 C_2)^2\right)^3 + \textrm{lower terms} \ .
\end{aligned}
\eeq

\bigskip

\subsection*{Acknowledgments}

\noindent
This work was partially supported by the Alexander von Humboldt Foundation,
Fondecyt grant 1171475 and by the Deutsche Forschungsgemeinschaft under grant LE 838/12.
This article is based upon work from COST Action MP1405 QSPACE,
supported by COST (European Cooperation in Science and Technology).
O.L.~is grateful for the warm hospitality at CECs and Universidad Austral de Chile,
where the main part of this work was done. F.C.~is also grateful for the warm hospitality
at Leibniz Universit\"at Hannover.

\bigskip

\appendix

\section{Potential-free frame}

\noindent
We display some relations with the potential-free frame for the $A_{N-1}$ model.
By conjugating the Hamiltonian one can trade the potential for a first-order derivative term,
\beq
\begin{aligned}
\widetilde{H}(g) \= \De^{-g}H(g)\,\De^{g}
&\=-\frac{1}{2}\sum^N_{i=1} \frac{\pa}{\pa q_i^2}
-\frac{\pi}{L}\,g \sum_{i<j}\cot\sfrac{\pi}{L}(q_i{-}q_j)
\Bigl(\frac{\pa}{\pa q_i}-\frac{\pa}{\pa q_j}\Bigr) \\[4pt]
&\= \frac{\pi^2}{L^2}\biggl[ 2\sum^N_{i=1} \Bigl(\frac{x_i\pa}{\pa x_i}\Bigr)^2
+2\,g\sum_{i<j}\frac{x_i{+}x_j}{x_i{-}x_j}\bigl(x_i\pa_i-x_j\pa_j\bigr)\biggr]\ .
\end{aligned}
\eeq

\section{Ring of Laurent polynomials}

\noindent
For convenience we describe the multiplication rule for the Laurent polynomials
\beq \label{Laurent}
\begin{aligned}
Q^{\pm}_{m_1,m_2}(y) &\=
y_1^{2m_1-m_2} y_2^{2m_2-m_1} y_3^{-m_1-m_2} \pm y_1^{2m_2-m_1} y_2^{2m_1-m_2} y_3^{-m_1-m_2}
+ \ \textrm{cyclic}\\[4pt]
&\= y_1^\al y_2^\be y_3^\ga + y_1^\be y_2^\ga y_3^\al + y_1^\ga y_2^\al y_3^\be
\pm y_1^\al y_2^\ga y_3^\be \pm y_1^\be y_2^\al y_3^\ga \pm y_1^\ga y_2^\be y_3^\al \\[4pt]
&\ =:\ [m_1,m_2]_\pm\ =:\ (\al,\be,\ga)_\pm
\qquad\with \al+\be+\ga=0\ .
\end{aligned}
\eeq
The relation between the parameters is
\beq
3m_1 = 2\al{+}\be = -\be{-}2\ga = \al{-}\ga  \und
3m_2 = 2\be{+}\al = -\al{-}2\ga = \be{-}\ga \ .
\eeq
To remove the redundancy of the labelling, we stipulate that
\beq
m_1 \ge m_2 \ge 0 \qquad\Leftrightarrow\qquad \al\ge\be\ge\ga\ .
\eeq
It is immediate that $[m_1,0]_-=[m_1,m_1]_-=0$.

The obvious multiplication ($\eps,\eb\in\{+,-\}$)
\beq
\begin{aligned}
& (\al,\be,\ga)_\eps \times (\ab,\bb,\gb)_\eb \= \\[4pt]
& \qquad\quad(\al{+}\ab,\be{+}\bb,\ga{+}\gb)_\ebe + (\al{+}\bb,\be{+}\gb,\ga{+}\ab)_\ebe + (\al{+}\gb,\be{+}\ab,\ga{+}\bb)_\ebe \\[4pt]
& \qquad +\eb\,(\al{+}\ab,\be{+}\gb,\ga{+}\bb)_\ebe +\eb\,(\al{+}\bb,\be{+}\ab,\ga{+}\gb)_\ebe +\eb\,(\al{+}\gb,\be{+}\bb,\ga{+}\ab)_\ebe 
\end{aligned}
\eeq
produces the law
\beq
\begin{aligned}
& [m_1,m_2]_\eps \times [n_1,n_2]_\eb \= \\[4pt]
& \qquad\quad [m_1{+}n_1,m_2{+}n_2]_\ebe +  [m_1{-}n_1{+}n_2,m_2{-}n_1]_\ebe+ [m_1{-}n_2,m_2{+}n_1{-}n_2]_\ebe\\[4pt]
& \qquad +\eb\,[m_1{+}n_1{-}n_2,m_2{-}n_2]_\ebe + \eb\,[m_1{+}n_2,m_2{+}n_1]_\ebe + \eb\,[m_1{-}n_1,m_2{-}n_1{+}n_2]_\ebe \ .
\end{aligned}
\eeq
Even assuming $m_1\ge n_1$ (without loss of generality),
the right-hand side may produce contributions $[k_1,k_2]_\pm$ with $k_1<k_2$ or with $k_2<0$,  which we outlawed.
However, it is easy to see that
\beq
[k_1,k_2]_\pm = \pm[k_2,k_1]_\pm \und [k_1,k_2]_\pm = \pm[k_1{-}k_2,-k_2]_\pm\ ,
\eeq
so one may employ the first relation in the first case and the second one in the second case to obtain an admissible result.
Some examples are
\beq
\begin{aligned}
[m_1,m_2]_\pm \times [0,0]_+ &\= 6\,[m_1,m_2]_\pm \ ,\\
[m_1,m_2]_\pm \times [1,0]_+ &\= 2\,[m_1{+}1,m_2]_\pm + 2\,[m_1,m_2{+}1]_\pm + 2\,[m_1{-}1,m_2{-}1]_\pm \ ,\\
[m_1,m_2]_\pm \times [1,1]_+ &\= 2\,[m_1{+}1,m_2{+}1]_\pm + 2\,[m_1,m_2{-}1]_\pm + 2\,[m_1{-}1,m_2]_\pm \ ,\\
[m_1,m_2]_\pm \times [2,0]_+ &\= 2\,[m_1{+}2,m_2]_\pm + 2\,[m_1,m_2{+}2]_\pm + 2\,[m_1{-}2,m_2{-}2]_\pm \ ,\\
[m_1,m_2]_\pm \times [2,2]_+ &\= 2\,[m_1{+}2,m_2{+}2]_\pm + 2\,[m_1,m_2{-}2]_\pm + 2\,[m_1{-}2,m_2]_\pm \ ,\\
[m_1,m_2]_\pm \times [2,1]_+ &\= [m_1{+}2,m_2{+}1]_\pm + [m_1{+}1,m_2{+}2]_\pm + [m_1{+}1,m_2{-}1]_\pm \\
& \ \ + [m_1{-}1,m_2{+}1]_\pm + [m_1{-}1,m_2{-}2]_\pm + [m_1{-}2,m_2{-}1]_\pm\ ,\\
[m_1,m_2]_\pm \times [2,1]_- &\= [m_1{+}2,m_2{+}1]_\mp - [m_1{+}1,m_2{+}2]_\mp - [m_1{+}1,m_2{-}1]_\mp \\
& \ \  + [m_1{-}1,m_2{+}1]_\mp 1+ [m_1{-}1,m_2{-}2]_\mp - [m_1{-}2,m_2{-}1]_\mp \ .
\end{aligned}
\eeq

\section{Wave functions for the $A_2$ model}

\noindent
The $A_2$ wave functions take the form
\beq
\langle x|n_1,n_2\rangle_g\ \equiv\ \Psi_{n_1,n_2}^{(g)}(x)
\= R_{n_1,n_2}^{(g)}(x) \,\De^g \= P_{n_1,n_2}^{(g)}(x)\,(x_1x_2x_3)^{-(n_1+n_2)/3}\,\De^g\ ,
\eeq
where $x=(x_1,x_2,x_3)$ and $P_{n_1,n_2}^{(g)}$ is a homogeneous
per\-mu\-ta\-tion-symmetric Jack polynomial of degree $n_1{+}n_2$ in~$x$.
Passing to the more convenient variables
\beq
y_i = x_i^{1/3}
\eeq
the rational function $R_{n_1,n_2}^{(g)}$ is a linear combination of symmetric basis functions
\beq
\begin{aligned}
&Q^+_{m_1,m_2}(y) 
\= y_1^{2m_1-m_2} y_2^{2m_2-m_1} y_3^{-m_1-m_2} + y_1^{2m_2-m_1} y_2^{2m_1-m_2} y_3^{-m_1-m_2}
+ \ \textrm{cyclic}\\[4pt]
&\textrm{with}\quad m_1{+}m_2=n_1{+}n_2{-}3\ell \for \ell=0,1,2,\ldots\ .
\end{aligned}
\eeq
The first few Laurent polynomials are
\begin{align} 
R_{0,0}&=Q^+_{0,0}=6 \ ,\\[3pt] \notag
R_{1,0}&=Q^+_{1,0}=2\,(y_1^2y_2^{-1}y_3^{-1}{+}y_2^2y_3^{-1}y_1^{-1}{+}y_3^2y_1^{-1}y_2^{-1}) \ ,\\[3pt] \notag
R_{1,1}&=Q^+_{1,1}=2\,(y_1\,y_2\,y_3^{-2}{+}y_2\,y_3\,y_1^{-2}{+}y_3\,y_1\,y_2^{-2}) \ ,\\[2pt]  \notag
R_{2,0}&=Q^+_{2,0}+\sfrac{2g}{g+1}Q^+_{1,1} \ ,\\ \notag
R_{2,1}&=Q^+_{2,1}+\sfrac{g}{2g+1}Q^+_{0,0} \ ,\\  \notag
R_{2,2}&=Q^+_{2,2}+\sfrac{2g}{g+1}Q^+_{1,0} \ ,\\  \notag
R_{3,0}&=Q^+_{3,0}+\sfrac{6g}{g+2}Q^+_{2,1}+\sfrac{2g^2}{(g+1)(g+2)}Q^+_{0,0} \ ,\\  \notag
R_{3,1}&=Q^+_{3,1}+\sfrac{g}{g+1}Q^+_{2,2}+\sfrac{g(5g+3)}{2(g+1)^2}Q^+_{1,0} \ ,\\  \notag
R_{3,2}&=Q^+_{3,2}+\sfrac{g}{g+1}Q^+_{2,0}+\sfrac{g(5g+3)}{2(g+1)^2}Q^+_{1,1} \ ,\\  \notag
R_{3,3}&=Q^+_{3,3}+\sfrac{6g}{g+2}Q^+_{2,1}+\sfrac{2g^2}{(g+1)(g+2)}Q^+_{0,0} \ ,\\  \notag
R_{4,0}&=Q^+_{4,0}+\sfrac{8g}{g+3}Q^+_{3,1}+\sfrac{6g(g+1)}{(g+2)(g+3)}Q^+_{2,2}
+\sfrac{12 g^2}{(g+2)(g+3)}Q^+_{1,0} \ ,\\  \notag
R_{4,1}&=Q^+_{4,1}+\sfrac{3g}{g+2}Q^+_{3,2}+\sfrac{g(7g+8)}{(g+2)(2g+3)}Q^+_{2,0}
+\sfrac{3g(4g+1)}{(g+2)(2g+3)}Q^+_{1,1} \ ,\\  \notag
R_{4,2}&= Q^+_{4,2}+\sfrac{g}{g+1} Q^+_{3,3}+\sfrac{g}{g+1} Q^+_{3,0}
+\sfrac{4g(4g+1)}{(g+1)(2g+3)}Q^+_{2,1}
+\sfrac{g(10g^2+7g+3)}{2(g+1)^2(2g+3)}Q^+_{0,0} \ ,\\  \notag
R_{4,3}&=Q^+_{4,3}+\sfrac{3g}{g+2}Q^+_{3,1}+\sfrac{g(7g+8)}{(g+2)(2g+3)}Q^+_{2,2}
+\sfrac{3g(4g+1)}{(g+2)(2g+3)}Q^+_{1,0} \ ,\\  \notag
R_{4,4}&=Q^+_{4,4}+\sfrac{8g}{g+3}Q^+_{3,2}+\sfrac{6g(g+1)}{(g+2)(g+3)}Q^+_{2,0}
+\sfrac{12g^2}{(g+2)(g+3)}Q^+_{1,1} \ ,\\ \notag
R_{5,0}&=Q^+_{5,0}+\sfrac{10g}{g+4}Q^+_{4,1}+\sfrac{20g(g+1)}{(g+3)(g+4)}Q^+_{3,2}
+\sfrac{20g^2}{(g+3)(g+4)}Q^+_{2,0}+\sfrac{30g^2(g+1)}{(g+2)(g+3)(g+4)}Q^+_{1,1}  \ ,\\ \notag
R_{5,1}&=Q^+_{5,1}+\sfrac{4g}{g+3}Q^+_{4,2}+\sfrac{3g(g+1)}{(g+2)(g+3)}Q^+_{3,3}
+\sfrac{3g(3g+5)}{(g+2)(g+3)}Q^+_{3,0}+\sfrac{2g(11g^2+20g+5)}{(g+2)^2(g+3)}Q^+_{2,1}
+\sfrac{6g^2(g+1)}{(g+2)^2(g+3)} Q^+_{0,0} \ .\\ \notag
\end{align}

{}From the paper of Lapointe and Vinet \cite{LaVi95}, one can see that the Jack polynomials 
can be also constructed in terms of modified Dunkl operators
\beq
{\cal D}_{i,\beta}(g)=x_i \Bigl( \pa_i +g \sum_{i\neq j} \frac{1-s_{ij}}{x_i-x_j} \Bigr)+\beta \,. 
\eeq
In terms of the combinations
\beq
\begin{aligned}
B_1(g) &\= x_1 {\cal D}_{1,g}(g)+x_2 {\cal D}_{2,g}(g)+x_3 {\cal D}_{3,g}(g) \ ,\\[4pt]
B_2(g) &\= x_1 x_2 {\cal D}_{1,g}(g){\cal D}_{1,2g}(g)+x_2 x_3{\cal D}_{2,g}(g){\cal D}_{3,2g}(g)+x_3 x_1 {\cal D}_{3,g}(g) {\cal D}_{1,2g}(g)
\end{aligned}
\eeq
the Jack polynomials are given by 
\beq
P_{n_1,n_2}^{(g)} \= B_2(g)^{n_2} B_1(g)^{n_1-n_2}\cdot 1\ .
\eeq
With the normalization
\beq
\Psi_{n_1,n_2}^{(g)}\=\frac{1}{(g)_{n_1-n_2} (g)_{n_2} (2g{+}n_1{-}n_2)_{n_2}}\, 
P_{n_1,n_2}^{(g)}\,(x_1x_2x_3)^{-(n_1+n_2)/3}\,\De^g
\eeq
employing the Pochhammer symbol $(a)_n$, the action of the intertwiner takes the form
\beq
M_3(g)\,\Psi_{n_1,n_2}^{(g)} \= n_2 (n_1{+}g) (n_1{-}n_2) \Psi_{n_1-2,n_2-1}^{(g+1)} \ .
\eeq
Clearly, the $n_2{=}0$ and $n_2{=}n_1$ states are annihilated by $M_3(g)$.

\section{Wave functions for the $G_2$ model}

\noindent
In the variables $y=(y_i)=(x_i^{1/3})$
the $G_2$ wave functions take the form
\beq
\langle y|n_1,n_2\rangle_{\gs,\gl}\ \equiv\ \Psi_{n_1,n_2}^{(\gs,\gl)}(y)
\= R_{n_1,n_2}^{(\gs,\gl)}(y) \,\Psi_0^{(\gs,\gl)}(y)
\eeq
with $\Psi_0^{(\gs,\gl)}$ given in (\ref{G2ground}) and~(\ref{G2Deltas})
and $R_{n_1,n_2}^{(\gs,\gl)}$ being a degree-zero rational function in~$y$.
It turns out that for $n_1{+}n_2$ not divisible by three $\Psi$ does not in general factorize
as $R$ times $\Psi_0$ in the ring of Laurent polynomials, so $R$ is of a more general class.
However, the full wave function~$\Psi$ can be expressed in terms of the
symmetric and antisymmetric basis polynomials
\beq 
\begin{aligned}
&Q^{\pm}_{m_1,m_2}(y) \=
y_1^{2m_1-m_2} y_2^{2m_2-m_1} y_3^{-m_1-m_2} \pm y_1^{2m_2-m_1} y_2^{2m_1-m_2} y_3^{-m_1-m_2}
+ \ \textrm{cyclic}\\[4pt]
&\textrm{with}\quad m_1{+}m_2=n_1{+}n_2{-}3\ell \for \ell=0,1,\ldots\ .
\end{aligned}
\eeq

Some low-lying factorizable states are listed below. 
Beyond $\Psi_{0,0}^{\gs,\gl}=Q^+_{0,0}\,\De_{\textrm{S}}^{\gs}\De_{\textrm{L}}^{\gl}$ we have
\beq
\begin{aligned}
\Psi_{1,0}^{g,0} &\= [Q^+_{1,0}{+}Q^+_{1,1}]\,\De_{\textrm{S}}^g \ , \\[4pt]
\Psi_{2,0}^{g,0} &\= \bigl([Q^+_{2,0}{+}Q^+_{2,2}]+\sfrac{2g}{g+1}[Q^+_{1,0}{+}Q^+_{1,1}]\bigr)\,\De_{\textrm{S}}^g\ , \\[4pt]
\Psi_{2,1}^{g,0} &\= \bigl(Q^+_{2,1}+\sfrac{g}{2g+1}Q^+_{0,0})\,\De_{\textrm{S}}^g\ , \\[4pt]
\Psi_{3,0}^{g,0} &\= \bigl([Q^+_{3,0}{+}Q^+_{3,3}]+\sfrac{12g}{g+2}Q^+_{2,1}+\sfrac{4g^2}{(g+1)(g+2)}Q^+_{0,0}\bigr)\,\De_{\textrm{S}}^g\ , \\[4pt]
\Psi_{3,1}^{g,0} &\= \bigl([Q^+_{3,1}{+}Q^+_{3,2}]+\sfrac{g}{g+1}[Q^+_{2,0}{+}Q^+_{2,2}]+\sfrac{g(5g+3)}{2(g+1)^2}[Q^+_{1,0}{+}Q^+_{1,1}]\bigr)\,\De_{\textrm{S}}^g\ , \\[4pt]
\Psi_{4,0}^{g,0} &\= \bigl([Q^+_{4,0}{+}Q^+_{4,4}]+\sfrac{8g}{g+3}[Q^+_{3,1}{+}Q^+_{3,2}]+\sfrac{6g(g+1)}{(g+2)(g+3)}[Q^+_{2,0}{+}Q^+_{2,2}]+\sfrac{12 g^2}{(g+2)(g+3)}[Q^+_{1,0}{+}Q^+_{1,1}]\bigr)\,\De_{\textrm{S}}^g\ , \\[4pt]
\Psi_{4,1}^{g,0} &\= \bigl([Q^+_{4,1}{+}Q^+_{4,3}]+\sfrac{3g}{g+2}[Q^+_{3,1}{+}Q^+_{3,2}]+\sfrac{g(7g+8)}{(g+2)(2g+3)}[Q^+_{2,0}{+}Q^+_{2,2}]+\sfrac{3g(4g+1)}{(g+2)(2g+3)}[Q^+_{1,0}{+}Q^+_{1,1}]\bigr)\,\De_{\textrm{S}}^g\ , \\[4pt]
\Psi_{4,2}^{g,0} &\= \bigl(Q^+_{4,2}+\sfrac{g}{g+1}[Q^+_{3,0}{+}Q^+_{3,3}]+\sfrac{4g(4g+1)}{(g+1)(2g+3)}Q^+_{2,1}+\sfrac{g(10 g^2+7 g+3)}{2(g+1)^2(2g+3)}Q^+_{0,0}\bigr)\,\De_{\textrm{S}}^g\ ,
\end{aligned}
\eeq
with \ $\De_{\textrm{S}}=Q^-_{2,1}$ \ and \ $\De_{\textrm{L}}=\sfrac12(Q^+_{3,0}-Q^+_{3,3})$. \\
In addition, one can infer that
\beq
\begin{aligned}
\Psi_{2,1}^{0,g} &\= Q^+_{2,1}\,\De_{\textrm{L}}^g \ ,\\[4pt]
\Psi_{3,0}^{0,g} &\= \bigl([Q^+_{3,0}{+}Q^+_{3,3}]+\sfrac{2g}{2g+1}Q^+_{0,0}\bigr)\,\De_{\textrm{L}}^g \ ,\\[4pt]
\Psi_{4,2}^{0,g} &\= \bigl(Q^+_{4,2}+\sfrac{2g}{g+1}Q^+_{2,1}\bigr)\,\De_{\textrm{L}}^g \ ,\\[4pt]
\Psi_{5,1}^{0,g} &\= \bigl([Q^+_{5,1}{+}Q^+_{5,4}]+\sfrac{2g}{g+1}Q^+_{4,2}+\sfrac{g(5g+3)}{(g+1)^2}Q^+_{2,1}\bigr)\,\De_{\textrm{L}}^g \ ,\\[4pt]
\Psi_{6,0}^{0,g} &\= \bigl([Q^+_{6,0}{+}Q^+_{6,6}]+\sfrac{4g}{g+1}Q^+_{6,3}+\sfrac{4g(4g+1)}{(g+1)(2g+3)}[Q^+_{3,0}{+}Q^+_{3,3}]+\sfrac{g(10g^2+7g+3)}{(g+1)^2(2g+3)}Q^+_{0,0}\bigr)\,\De_{\textrm{L}}^g \ ,\\[4pt]
\Psi_{6,3}^{0,g} &\= \bigl(Q^+_{6,3}+\sfrac{3g}{g+2}[Q^+_{3,0}{+}Q^+_{3,3}]+\sfrac{2g^2}{(g+1)(g+2)}Q^+_{0,0}\bigr)\,\De_{\textrm{L}}^g \ ,\\[4pt]
\Psi_{7,2}^{0,g} &\= \bigl([Q^+_{7,2}{+}Q^+_{7,5}]+\sfrac{3g}{g+2}[Q^+_{5,1}{+}Q^+_{5,4}]+\sfrac{2g(7g+8)}{(g+2)(2g+3)}Q^+_{4,2}+\sfrac{6g(4g+1)}{(g+2)(2g+3)}Q^+_{2,1}\bigr)\,\De_{\textrm{L}}^g \ .
\end{aligned}
\eeq
For $\gs,\gl\in\{0,1\}$, the model is free, so the states take a very simple form:
\beq
\begin{aligned}
\Psi^{0,0}_{n_1,n_2} &\= Q^+_{n_1,n_2} + Q^+_{n_1,n_1-n_2} \ ,\\[4pt]
\Psi^{1,0}_{n_1,n_2} &\= Q^-_{n_1+2,n_2+1} + Q^-_{n_1+2,n_1-n_2+1} \ ,\\[4pt]
\Psi^{0,1}_{n_1,n_2} &\= Q^+_{n_1+3,n_2} - Q^+_{n_1+3,n_1-n_2+3} \ ,\\[4pt]
\Psi^{1,1}_{n_1,n_2} &\= Q^-_{n_1+5,n_2+1} - Q^-_{n_1+5,n_1-n_2+4} \ .
\end{aligned}
\eeq
In the last two lines, $\De_{\textrm{L}}$ can only be factored off in case $n_1{+}n_2$ is a multiple of three.

\section{Coefficients of the conserved charge $J_6$}

\noindent
In order to express the coefficients $c_{s_1s_2s_3}$ in (\ref{Jsix}) we introduce the combinations 
\begin{align}
\Omega_{ij}=\frac{x_i\,x_j}{(x_i{-}x_j)^2} \und
\Upsilon_{k}= \frac{x_k^2\,x_i \,x_j}{(x_k^2{-}x_i x_j)^2}
\end{align}
as well as
\begin{align}
\widetilde{\Omega}^n_{ij}=\frac{x_i{+}x_j}{x_i{-}x_j}\,\Omega_{ij}^n \und
\widetilde{\Upsilon}_{k}^n= \frac{x_k^2{+}x_i x_j}{x_k^2{-}x_i x_j}\,\Upsilon_{k}^n\, ,
\end{align}
where $\{i,j,k\}=\{1,2,3\}$ and $n$ is a positive integer.
In this way the coefficients read

{\small
\beq
\begin{aligned}
c_{400} &\= -12\,\gs(\gs{-}1)(3 \Omega_{23}{+}4\Omega_{31})-36\,\gl(\gl{-}1)(2 \Upsilon_{1}{+}4\Upsilon_{3}) -75 \gls{-}84 \gl\gs{-}28 \gss\, , \\[6pt]
c_{310} &\= -12\,\gs(\gs{-}1)(7 \Omega_{12}{+}4 \Omega_{23}{+}9\Omega_{31})-36\,\gl(\gl{-}1)(8 \Upsilon_{1}{+}5 \Upsilon_{2}{+}11\Upsilon_{3}) \\
 &\qquad\! -4(69 \gls{+}60 \gl\gs{+}20 \gss)\, ,\\[6pt]
c_{220} &\=  -12\,\gs(\gs{-}1)(3 \Omega_{12}{+}3\Omega_{31})-36\,\gl(\gl{-}1)(5 \Upsilon_{2}{-}2\Upsilon_{3})-3(15 \gls{+}24 \gl\gs{+}8 \gss)\, , \\[6pt]
c_{211} &\= -12\,\gs(\gs{-}1)(3 \Omega_{23}{-}9\Omega_{31})+36\,\gl(\gl{-}1)(5 \Upsilon_{1}{+}\Upsilon_{3})+24(3 \gls{+}3 \gl\gs{+}\gss)\, ,
\end{aligned}
\eeq
\beq
\begin{aligned}
c_{300} &\=-54\,\gs(\gs{-}1)\widetilde{\Omega}_{31}+18\,\gl(\gl{-}1)(8\widetilde{\Upsilon}_{1}{-}17\widetilde{\Upsilon}_{3})\, ,\\[6pt]
c_{210} &\=54\,\gs(\gs{-}1)(\widetilde{\Omega}_{12}{-}4\widetilde{\Omega}_{31})+18\,\gl(\gl{-}1)(48\widetilde{\Upsilon}_{1}{+}6 \widetilde{\Upsilon}_{2}{-}27\widetilde{\Upsilon}_{3})\, ,
\qquad\qquad\qquad\quad\ {}\\[6pt]
c_{111} &\=  -18^2\,\gl(\gl{-}1)\widetilde{\Upsilon}_{3}\, ,
\end{aligned}
\eeq
\vspace{1mm}
\beq
\begin{aligned}
c_{200} &\= 36\,\gs(\gs{-}1)(2\gs{+}3\gl)^2(\sfrac{1}{2} {\Omega}_{23}{+}{\Omega}_{31})+108\,\gl(\gl{-}1)(2\gs{+}3\gl)^2(\sfrac{1}{2} {\Upsilon}_{1}{+}{\Upsilon}_{3})\\
&\ -54\, \gs(\gs{-}1) {\Omega}_{31}-36\,\gl(\gl{-}1)(2 \Upsilon_{1}{+}\sfrac{47}{2}{\Upsilon}_{3})+54\,\gs(\gs{-}1)(\gs{+}2)(\gs{-}3) {\Omega}_{31}^2\\
&\ +36\,(\gl{+}1)\gl(\gl{-}1)(\gl{-}2) (6{\Upsilon}_{1}^2{+}\sfrac{51}{2} {\Upsilon}_{3}^2)-36\cdot 90\,\gl(\gl{-}1) {\Upsilon}_{3}^2\\
&\ +54\, \gs^2(\gs{-}1)^2(4\Omega_{23}\Omega_{31}{+}\Omega_{31}\Omega_{12})+108\, \gs(\gs{-}1)\gl(\gl{-}1)(-4\Omega_{23}\Omega_{31}+\Omega_{31}\Omega_{12})\\
&\ +54\, \gl^2(\gl{-}1)^2(-{\Upsilon}_{2}{\Upsilon}_{3}{+}16{\Upsilon}_{3}{\Upsilon}_{1})+  18 \left(2 \gs^4{+}12 \gl \gs^3{+}30 \gl^2 \gs^2{+}36 \gl^3 \gs{+}15 \gl^4\right) \\
&\ +36\, \gs(\gs{-}1)\gl(\gl{-}1)(17\Omega_{12}\Upsilon_{2}{+}14\Omega_{23}\Upsilon_{3}{+}8\Omega_{31}\Upsilon_{1}{+}10\Upsilon_{1}\Omega_{23}{+}5
\Upsilon_{3}\Omega_{12})\, , 
\end{aligned}
\eeq
\vspace{1mm}
\beq
\begin{aligned}
c_{110} &\= 36\,\gs(\gs{-}1)(8\gs^2{+}24\gs \gl{+}27 \gl^2)(\sfrac{1}{2} {\Omega}_{12}{+}{\Omega}_{31}) \\
&\ +108\, \gl(\gl{-}1)(8\gs^2{+}24\gs \gl{+}27 \gl^2)({\Upsilon}_{2}{+}\sfrac{1}{2} {\Upsilon}_{3}) \\
&\ -108\, \gs(\gs{-}1)(\sfrac{1}{2} {\Omega}_{12}{+}2{\Omega}_{31})-36\,\gl(\gl{-}1)(50 \Upsilon_{2}{+}\sfrac{11}{2}{\Upsilon}_{3})\\
&\ +108\,\gs(\gs{-}1)(\gs{+}2)(\gs{-}3)(\sfrac{1}{2} {\Omega}_{12}^2{+}2{\Omega}_{31}^2)\\
&\ +108\, \gl^2(\gl{-}1)^2(14{\Upsilon}_{2}^2{+}\sfrac{5}{2} {\Upsilon}_{3}^2)-108\,\gl(\gl{-}1) (100{\Upsilon}_{2}^2{+}11{\Upsilon}_{3}^2)\\
&\ +54\, \gs^2(\gs{-}1)^2(\Omega_{23}\Omega_{31}{+}6\Omega_{31}\Omega_{12})+54\, \gs(\gs{-}1)\gl(\gl{-}1)(10\Omega_{23}\Omega_{31}-4\Omega_{31}\Omega_{12})\\
&\ +54\, \gl^2(\gl{-}1)^2(19{\Upsilon}_{1}{\Upsilon}_{2}{+}74{\Upsilon}_{3}{\Upsilon}_{1})+18 \left(4 \gs^4+24 \gl \gs^3+60 \gl^2 \gs^2+72 \gl^3 \gs+39 \gl^4\right)\\
&\ +36\, \gs(\gs{-}1)\gl(\gl{-}1)(25\Omega_{12}\Upsilon_{2}{+}13\Omega_{23}\Upsilon_{3}{+}31\Omega_{31}\Upsilon_{1}{+}16\Upsilon_{2}\Omega_{31}{+}23\Upsilon_{3}\Omega_{12})\, , 
\end{aligned}
\eeq
\vspace{1mm}
\beq
\begin{aligned}
c_{100} &\= -162\,\gs(\gs{-}1) \gl^2\, \widetilde{\Omega}_{31}+54\,\gl(\gl{-}1)(9 \gl^2{-}8)(\widetilde{\Upsilon}_{1}-\widetilde{\Upsilon}_{3}) \\
&\ + 36 \cdot 18(\gl(\gl{-}1){-}8) \gl(\gl{-}1)(\widetilde{\Upsilon}_{1}^2-\widetilde{\Upsilon}_{3}^2) \\
&\ -162\gs(\gs{-}1)(\gs(\gs{-}1){-}6\gl(\gl{-}1)) (\widetilde{\Omega}_{12}\Omega_{23}+\widetilde{\Omega}_{31}\Omega_{12}) \\
&\ +162 \gl^2(\gl{-}1)^2(8\widetilde{\Upsilon}_{1}\Upsilon_{2}-13\widetilde{\Upsilon}_{2}\Upsilon_{3}+5\widetilde{\Upsilon}_{3}\Upsilon_{1}) \\
&\ -54 \gs(\gs{-}1) \gl(\gl{-}1)(3\widetilde{\Omega}_{12}\Upsilon_{2}{-}6\widetilde{\Omega}_{23}\Upsilon_{3}{+}3\widetilde{\Omega}_{31}\Upsilon_{1}+8\widetilde{\Upsilon}_{1}\Omega_{23}{-}8\widetilde{\Upsilon}_{3}\Omega_{12}) \qquad\qquad\ {}\\
&\ -54 \gs(\gs{-}1) \gl(\gl{-}1)({\Omega}_{12}\widetilde{\Upsilon}_{2}+7{\Omega}_{23}\widetilde{\Upsilon}_{3}-8{\Omega}_{31}\widetilde{\Upsilon}_{1}-12{\Upsilon}_{3}\widetilde{\Omega}_{12})\, 
\end{aligned}
\eeq
\vspace{1mm}
\beq
\begin{aligned}
c_{000} &\= 27 \gl^6+81\, \gs(\gs{-}1) \gl^2 \left(2 \gl^2-1\right)\Omega_{12} +81\, \gl(\gl{-}1)(6 \gl^4{-}9 \gl^2{+}4)  \left(2 \gl^2-1\right)\Upsilon_{1} \\
&\ + 81\,  \gl^2\, \gs(\gs{-}1)(\gs{+}2)(\gs{-}3)\Omega_{12}^2+81\,  \gl(\gl{-}1)(\gl{+}2) (\gl{-}3) \left(9 \gl^2{-}20\right) \Upsilon_{1}^2 \\
&\ +324\,  \gl(\gl{-}1)(\gl{+}2)(\gl{-}3)(\gl{+}4)(\gl{-}5)\Upsilon_{1}^3 \\
&\ +162\, \gl \gs^2 (\gs{-}1)^2 \, \Omega_{12} \Omega_{23} +162\, \gl^2 (\gl{-}1)^2 (9 \gl^2{+}9\gs(\gs{-}1){-}8) \,\Upsilon_{1}\Upsilon_{2} \\
&\  +324 \,\gl \gs(\gs{-}1) (\gl{-}1)  \Omega_{23} ( (\sfrac{3}{2}\gl{-1})\Upsilon_{1}+(3\gl^2{-1})\Upsilon_{3}) \\
&\ -324 \, \gl \gs^2 (\gs{-}1)^2 (\gl{-}1) \, \Omega_{12}^2 \Omega_{23}+1944\, \gl^2(\gl{-}1)^2  (\gl{+}2)(\gl{-}3) \Upsilon_{1}^2 \Upsilon_{2} \\
&\ +324 \,\gl \gs(\gs{-}1) (\gl{-}1)   \Omega_{12}^2(\gs(\gs{-}1) \Upsilon_{1}+(\gs{+}2)(\gs{-}3) \Upsilon_{3}) \\
&\ +648 \, \gl  \gs (\gs{-}1) (\gl{-}1)(\gl{+}2)(\gl{-}3)  \Upsilon_{1}^2(\Omega_{12}-\Omega_{23}) \\
&\ +162 \, \gl^2 \gs (\gs{-}1)(\gl{-}1)^2 \, \Omega_{12} \Upsilon_{1}(13\Upsilon_{2}+14\Upsilon_{3}) \\
&\ -324  \,\gl \gs(\gs{-}1) (\gl{-}1)(2\gs(\gs{-}1){+}3 \gl(\gl{-}1)+8) \Omega_{12}\Omega_{23}\Upsilon_{1} \\
&\ -162  \,\gl \gs(\gs{-}1) (\gl{-}1)(3 \gl(\gl{-}1)-16) \Omega_{12}\Omega_{31}\Upsilon_{1} \\
&\ +648\, \gl^2(\gl{-}1)^2 (\gl^2{-}\gl{+}4)\Upsilon_{1}\Upsilon_{2}\Upsilon_{3}\, .
\end{aligned}
\eeq
}

%

%\newpage
%\bigskip

{\small
%\phantom{.}\bigskip

}


\begin{thebibliography}{99}
\addtolength{\itemsep}{-3pt}

\bibitem{Fring12-rev}
A. Fring,\\
{\it ${\cal PT}$-symmetric deformations of integrable models},\\
Phil. Trans. Roy. Soc. Lond. A {\bf 371} (2013) 20120046 
\href{https://arxiv.org/abs/1204.2291}{{\tt [arXiv:1204.2291[hep-th]]}}.

\bibitem{FrZn08}
A. Fring, M. Znojil,\\  
{\it ${\cal PT}$-symmetric deformations of Calogero models},\\
J. Phys. A {\bf 41} (2008) 194010 
\href{https://arxiv.org/abs/0802.0624}{{\tt [arXiv:0802.0624[hep-th]]}}.

\bibitem{fringsmith1} 
A. Fring, M. Smith, \\
{\it Antilinear deformations of Coxeter groups, an application to Calogero models},\\
J. Phys. A {\bf 43} (2010) 325201 
\href{https://arxiv.org/abs/1004.0916}{{\tt [arXiv:1004.0916[hep-th]]}}.
  
\bibitem{fringsmith2} 
A. Fring, M. Smith, \\
{\it ${\cal PT}$ invariant complex $E_8$ root spaces},\\
Int. J. Theor. Phys. {\bf 50} (2011) 974 
\href{https://arxiv.org/abs/1010.2218}{{\tt [arXiv:1010.2218[math-ph]]}}.
  
\bibitem{fringsmith3} 
A. Fring, M. Smith, \\
{\it Non-Hermitian multi-particle systems from complex root spaces},\\
J. Phys. A {\bf 45} (2012) 085203 
\href{https://arxiv.org/abs/1108.1719}{{\tt [arXiv:1108.1719[hep-th]]}}.

\bibitem{CoLe17}
F. Correa, O. Lechtenfeld,\\
{\it $\cal PT$ deformation of angular Calogero models},\\
JHEP {\bf 1711} (2017) 122
\href{https://arxiv.org/abs/1705.05425}{{\tt [arXiv:1705.05425[hep-th]]}}.

\bibitem{CoLePl13}
F. Correa, O. Lechtenfeld, M. Plyushchay,\\
{\it Nonlinear supersymmetry in the quantum Calogero model,}\\
JHEP {\bf 1404} (2014) 151
\href{https://arxiv.org/abs/1312.5749}{{\tt [arXiv:1312.5749[hep-th]]}}.

\bibitem{LaVi95}
L. Lapointe, L. Vinet,\\
{\it Exact operator solution of the Calogero--Sutherland model},\\
Commun. Math. Phys. {\bf178} (1996) 425
\href{https://arxiv.org/abs/q-alg/9509003}{{\tt [arXiv:q-alg/9509003]}}.

\bibitem{PeRaZa98}
A.M. Perelomov, E. Ragoucy, Ph. Zaugg,\\
{\it Explicit solution of the quantum three-body Calogero--Sutherland model},\\
J. Phys. A: Math. Gen. {\bf 31} (1998) L559
\href{https://arxiv.org/abs/hep-th/9805149}{{\tt [arXiv:hep-th/9805149]}}.

\bibitem{GaLoPe01}
W. Garc\'ia Fuertes, M. Lorente, A.M. Perelomov,\\
{\it An elementary construction of lowering and raising operators for 
the trigonometric Calogero--Sutherland model},\\
J. Phys. A: Math. Gen. {\bf 34} (2001) 10963
\href{https://arxiv.org/abs/math-ph/0110038}{{\tt [arXiv:math-ph/0110038]}}.

\bibitem{Qu95}
C.~Quesne,\\
{\it Exchange operators and extended Heisenberg algebra for the three-body Calogero--Marchioro--Wolfes problem},\\
Mod. Phys. Lett. A {\bf 10} (1995) 1323
\href{https://arxiv.org/abs/hep-th/9505071}{{\tt [arXiv:hep-th/9505071]}}.

\bibitem{Qu96}
C.~Quesne,\\
{\it Three-body generalization of the Sutherland model with internal degrees of freedom},\\
Europhys. Lett. {\bf 35} (1996) 407
\href{https://arxiv.org/abs/hep-th/9607035}{{\tt [arXiv:hep-th/9607035]}}.

\end{thebibliography}
\end{document}